\documentstyle[preprint,aps,epsf]{revtex}
\newcommand{\postscript}[2] {\setlength{\epsfxsize}{#2\hsize}
\centerline{\epsfbox{#1}}}
\begin{document}
\title{Preprint MPI~H-V3-2001 \\
Renormalization of an effective Light-Cone QCD-inspired \\
       theory for the Pion and other Mesons}
\author{T. Frederico}
\address{Dep.de F\'\i sica, 
         Instituto Tecnol\'ogico de Aeron\'autica, 
         Centro T\'ecnico Aeroespacial, \\
         12.228-900 S\~ao Jos\'e dos Campos, S\~ao Paulo, Brazil 
\\
\text{and} }
%
%\and
% 
\author{Hans-Christian Pauli}
\address{Max-Planck Institut f\"ur Kernphysik, D-69029 Heidelberg, Germany}
\date{Draft 21 March 2001}
\maketitle
\begin{abstract}
    The renormalization of the effective QCD-Hamiltonian
    theory for the quark-antiquark channel is performed
    in terms of a renormalized or fixed-point Hamiltonian 
    that leads to subtracted dynamical equations. 
    The fixed point-Hamiltonian brings the renormalization 
    conditions as well as  the counterterms that render the 
    theory finite. The approach is renormalization group invariant. 
    The parameters of the renormalized effective  QCD-Hamiltonian 
    comes from the pion mass and radius, 
    for a given constituent quark mass. 
    The 1s and excited 2s states of $\overline u q$ are 
    calculated as a function of the mass of the quark $q$ 
    being s, c or b, and compared to the experimental values.
\end{abstract}


\section{Introduction}

The effective mass operator equation for the lowest Light-Front Fock-state 
component of a bound system of a constituent quark and antiquark of masses
$m_1$ and $m_2$,
obtained in the effective one-gluon-exchange interaction approximation,
which is our starting point, was 
reviewed in Ref.\cite{pauli0}. The breakthrough to simplify this
equation in the spin zero channel was achieved in \cite{pauli2}
 by formulating the $\uparrow\downarrow$-model, which is reduced to
\begin{eqnarray}
M^2\psi (x,{\vec k_\perp})&=&
\left[\frac{{\vec k_\perp}^2+m^2_1}{x}+\frac{{\vec k_\perp}^2+m^2_2}{1-x}
\right]\psi (x,{\vec k_\perp})
\nonumber \\
&-&\int \frac{dx' d{\vec k'_\perp}\theta(x')\theta (1-x')}
{\sqrt{x(1-x)x'(1-x')}}
\left(\frac{4m_1m_2}{3\pi^2}\frac{\alpha}{Q^2}-\lambda\right)
\psi (x',{\vec k'_\perp}) \ ,
\label{p1}
\end{eqnarray}
where $M$ is the mass of the bound-state and $\psi$ is the
projection of the light-front wave-function in the quark-antiquark Fock-state.
The mean four-momentum transfer is  \cite{pauli1}
\begin{eqnarray}
Q^2(x,\vec k_\perp,x',\vec k'_\perp)=-\frac12
\left[(k_1-k'_1)^2-(k_2-k'_2)^2\right] \ .
\label{q}
\end{eqnarray}
The coupling constant $\alpha$ for the Coulomb-like potential 
and $\lambda$ is the bare coupling constant of the Dirac-delta 
hyperfine interaction. For the purposes of this work, we leave
out the energy transfer in Eq.(\ref{q}), that leads to aditional 
singularities for the tridimensional momentum going to infinity,
that we are not going to treat here. The effective mass operator 
of Eq.(\ref{p1}) acting in the quark-antiquark sector,
has also been obtained by  iterated resolvents, which were 
derived in \cite{pauli3} and presented in greater detail
in \cite{pauli4}, that allows to express systematically
the higher Fock-state components of the wave-function in 
functionals of the lower ones. In this way, the higher Fock-state
components can be retrieved from the $q\overline q$ projection,
and the full complexity of the QCD theory is in principle 
described by the effective Hamiltonian acting in the lowest Fock-state
component\cite{pauli3}.

For convenience, the Sawicki tranformation,
first derived for equal masses \cite{saw1}  
and consistently
formulated for unequal masses in Ref. \cite{pauli1}, 
is applied to Eq.(\ref{p1}) 
which allows to rewrite it in the instant-form momentum basis.
It is useful in this case since the momentum transfer  
is approximated by a rotational invariant form given by Eq.(\ref{q}):
\begin{eqnarray}
x(k_z)=\frac{(E_1+k_z)}{E_1+E_2} \ ,
\label{xkz}
\end{eqnarray}
and the Jacobian of the transformation of $(x,{\vec k_\perp})$ 
to $\vec k$ is:
\begin{eqnarray}
dx d\vec k_\perp= \frac{x(1-x)}{m_r A(k)}d\vec k \ ,
\end{eqnarray}
with the dimensionless function 
\begin{eqnarray}
A(k)=\frac{1}{m_r}\frac{E_1 E_2}{E_1+E_2} \ ,
\label{phsp}
\end{eqnarray}
and the reduced mass $m_r=m_1m_2/(m_1+m_2)$. The individual
energies are $E_i=\sqrt{m_i^2+k^2}$ ($i$=1,2) and $k\equiv|\vec k|$.

The mass operator equation in instant form momentum variables is given by:
\begin{eqnarray}
M^2\varphi(\vec k)=
\left[E_1+E_2\right]^2\varphi(\vec k)-\int d\vec k'
\left(\frac{4m_s}{3\pi^2}\frac{\alpha}{\sqrt{ A(k)A(k')}Q^2}-
\frac{\lambda}{m_r\sqrt{ A(k)A(k')}}\right)
\varphi (\vec k') \ ,
\label{mass1}
\end{eqnarray}
where $ m_s=m_1+m_2$, the phase-space factor is included in 
the factor $1/\sqrt{ A(k)A(k')}$ and 
$\sqrt{x(1-x)}\psi (x,\vec k_\perp)=\sqrt{A(k)}\varphi(\vec k)$.

The mass operator equation (\ref{mass1}) needs
to be regularized and renormalized in order to give physical
results. In Ref. \cite{pauli2}, the delta function was smeared out
to a Yukawa form, regularizing Eq.(\ref{mass1}) and the
parameters found from  the pion mass and radius. In principle
the size parameter in momentum space of the Yukawa potential should
be let to infinity while the physical input is kept constant. To make 
vanish the dependence on the size parameter
is a non trivial task, which is the purpose of this work. 
Here we obtain a renormalized form of the equation for the bound state 
mass, which is {\it i}) invariant under renormalization group 
transformations, {\it ii}) the physical input is given
by the pion mass and radius {\it iii}) no regularization parameter. 

We are going to apply the recent renormalization techniques developed 
in the context non-relativistic Hamiltonian theory \cite{t1,t2,t3} to
Eq.(\ref{mass1}), since it defines an effective Hamiltonian for the
quark-antiquark dynamics. The renormalized T-matrix is  the solution
of a subtracted scattering equation, which the physical input 
is given by the T-matrix at some reference scale $\mu$. The scheme 
is invariant under the change of the arbitray scale $\mu$ and 
consequently the inhomogeneous term of the subtracted scattering 
equation satisfies a renormalization
group equation, which expresses the matching of the theories at the
scales $\mu$ and $\mu+d\mu$ \cite{t2}. We will find
the mass of the bound state from the pole of the renormalized scattering
matrix defined from the mass operator Eq.(\ref{mass1}). 

This work is organized as follows. In sec.II, the  operators
for the Coulomb-like and for the singular interactions appearing in
the mass operator are defined, and the Lippman-Schwinger 
equation for the T-matrix related to the given  mass operator
is written. In sec. III, we show how to renormalized the effective
theory defined by the mass operator in sec.II through the definition of
a renormalized or fixed point singular interaction. This procedure
is equivalent to use subtracted scattering equations. We discuss the
renormalization group invariance of the method. The explicit form
of the renormalized T-matriz is obtained (appendix A) and the 
physical input is the pion mass which determines the excited states
as well as the mass of the other mesons. The physical observables
of the renormalized effective theory does not depend on the subtraction
point. In sect.IV, the calculation of the pion charge radius is
discussed and although we have simplified the spin dependence in
the dynamical equation it is important in the evaluation of the radius.
 For this purpose we have used an effective pseudo-scalar
Lagrangian to  construct the spin part of the pion wave-function, 
and turning off the Coulomb-like interaction we retrieve a well
knwon result valid in the soft pion limit\cite{tarr}. To gain insight
we also write down the expressions where the quark spin is neglected.
The numerical results are presented in sec. V. We have 
solved a nonrelativistic  example for a Coulomb plus a Dirac-delta 
interaction compared to a model in which the Dirac-delta is substituted
by a Yukawa. We also show how this effective theory, including a
Dirac-delta, can mimic a finite range theory, calibrated to the 
pion mass. Then, we present results for the pion charge radius
and for the mass of the excited state, that we idendified with the
isovector vector mesons. Finally, in sec. VI, we present our conclusions.

\section{Definitions}

For our puposes
it is convenient to work in an operator form of Eq.({\ref{mass1}):
\begin{eqnarray}
\left( M_0^2+V+V^\delta \right)|\varphi >= M^2 |\varphi >\ ,
\label{mass2}
\end{eqnarray}
and the free mass operator $M_0$ is the sum of the energies of 
quark 1 and 2, $V$  the Coulomb-like potential and $V^\delta$ is
the Dirac-delta interaction in the non-relativistic limit. 
The matrix elements of these operators are given by:
\begin{eqnarray}
<\vec k|V|\vec{k'}>&=&-\frac{4m_s}{3\pi^2}\frac{\alpha}{\sqrt{ A(k)}
Q^2\sqrt{A(k')}}
\ ,
\label{mecoul}
\end{eqnarray}
and the short-range singular interaction
\begin{eqnarray}
   <\vec k|V^\delta|\vec{k'}>=
   <\vec k|\chi> \frac{\lambda}{m_r}<\chi|\vec{k'}> = 
   \frac{\lambda}{m_r}
   \frac{1}{\sqrt{A(k)}}
   \frac{1}{\sqrt{A(k')}} .
\label{mesing} \end{eqnarray}
The phase-space factor $A(k)$ is defined by Eq.(\ref{phsp}), 
the square momentum transfer $Q^2$ comes from Eq.(\ref{q}).
For convenience in the formal manipulations of the next section, the 
form-factor of the separable singular interaction is
introduced and defined by $<\vec k|\chi>=1/\sqrt{A(k)}$.

The T-matrix is obtained from the Lippman-Schwinger equation:
\begin{eqnarray}
T(M^2)=V+V^\delta + (V+V^\delta)G^{(+)}_0(M^2)T(M^2) \ ,
\label{t1}
\end{eqnarray}
where $M$ is the mass of the scattering state and the Green's function
with outgoing wave boundary condition is 
\begin{eqnarray}
G^{(+)}_0(M^2)=\frac{1}{M^2-M^2_0+i\varepsilon}\ .
\label{g0}
\end{eqnarray}

In the next section we will obtain a renormalized form of 
$T(M^2)$.

\section{ Renormalization of the Effective Theory}

The ideas that have been developed  in Refs. \cite{t1,t2,t3} 
to construct a renormalized T-matrix in non-relativistic Hamiltonian
theory, can be applied in the case where the two-body interaction can be 
splited in a regular potential of finite range $(V)$ and a Dirac-delta 
singularity as in Eq.(\ref{t1}). In this case, the physical information 
at the subtraction point will be introduced through the renormalized 
mass operator as well as all the counterterms that render finite
the T-matrix equation (\ref{t1}). The physical information at
the subtraction point is the T-matrix corresponding to the Dirac-delta 
interaction. This renormalization approach has been
applied in the two-nucleon system to calculate the
T-matrix of the one-pion-exchange potential, without the
necessity of regularization or form-factors\cite{t1}.

The renormalized interaction is given by:
\begin{eqnarray}
V_{\cal R}&=& V+V^\delta_{\cal R} \ ,
\label{vfren}
\end{eqnarray}
where the renormalized Dirac-delta interaction is given by:
\begin{eqnarray}
V^\delta_{\cal R} &=&
\frac{1}{1+T^\delta_{\cal R}(\mu^2) G^{(+)}_0(\mu^2)}
T^\delta _{\cal R}(\mu^2)
\nonumber \\ 
&=&T^\delta _{\cal R}(\mu^2)\frac{1}{1+ G^{(+)}_0(\mu^2)
T^\delta _{\cal R}(\mu^2)}
\nonumber \\ 
&=&
T^\delta_{\cal R}(\mu^2)\sum^\infty_{n=0} 
\left[ -G_0(\mu^2)
T^\delta_{\cal R}(\mu^2)\right]^n 
\ ,
\label{vfren1}
\end{eqnarray}
and $T^\delta_{\cal R}(\mu^2)$ is the renormalized T-matrix of 
the Dirac-delta interaction, with matrix elements given by:
\begin{eqnarray}
<\vec p|T^\delta_{\cal R}(\mu^2)|\vec q>=
<\vec p|\chi>\lambda_{\cal R}(\mu^2) <\chi|\vec q>\ ,
\label{medelta}
\end{eqnarray}
where $\lambda_{\cal R}(\mu^2)$ is the renormalized strength of the
Dirac-delta interaction at the mass scale $\mu^2$.
In the non-relativistic limit the form-factor $\chi(q)=1$,
and the renormalized interaction becomes the Dirac-delta. 
 The scattering equation
with the renormalized interaction appears in a subtracted form \cite{t1,t2,t3}
after a little rearrangement of terms, in which
all the divergent momentum integrals are removed and it is written as
\begin{eqnarray}
   T_{\cal R}(M^2)&=&T_{\cal R}(\mu^2)+T_{\cal R}(\mu^2) 
   \left( G^{(+)}_0(M^2)- G^{(+)}_0(\mu^2)
   \right)T_{\cal R}(M^2) 
.\label{tvren7} \end{eqnarray} 
It is presented here in a general way, where we have dropped the  $\delta$ 
superscript, just to remind the reader that for a regular potential 
Eq.(\ref{tvren7}) is  completly equivalent to the traditional 
Lippman-Schwinger scattering equation.

The renormalized interaction is independent
of the subtraction point, i.e., 
the physics expressed by the renormalized
interaction is invariant by changes in the arbitrary renormalization point,
this physical requirement is given by:
\begin{eqnarray}
\frac{d}{d\mu^2}V^\delta_{\cal R}=0 \ ;
\label{dv}
\end{eqnarray}
qualifying the interaction as the fixed-point of Eq.(\ref{dv}),
which implies that the T-matrix found from the solution of
\begin{eqnarray}
T_{\cal R}(M^2)=V+ V_{\cal R}^\delta 
+\left(V+ V_{\cal R}^\delta \right) G^{(+)}_0(M^2)T_{\cal R}(M^2)) \ ,
\label{trenv}
\end{eqnarray} 
is invariant under dislocations of the subtraction point. Consequently, the
renormalized coupling constant of the Dirac-delta interaction changes as
the subtraction point moves, according to the Callan-Symanzik equation
\begin{eqnarray}
\frac{d}{d \mu^2} T^\delta_{\cal R}(\mu^2)=-T^\delta_{\cal R}(\mu^2)
\frac{1}{\left(\mu^2+i\varepsilon-M^2_0\right)^2}T^\delta _{\cal R}(\mu^2) \ ,
\label{tren7}
\end{eqnarray} 
obtained from Eq.(\ref{dv}).

Although the renormalized interaction is not well defined 
for singular interactions, the resulting T-matrix obtained by solving
Eq.(\ref{trenv}) is finite. This gives a posteriori justification
for the formal manipulations used  in (\ref{vfren1})  and (\ref{trenv}).
The  sum in the expression of the renormalized interaction (\ref{vfren1})
explicits all the  counterterms which exactly cancels the infinities in the
momentum integrals of the scattering equation (\ref{trenv}), 
while introducting the physical information through the value of the 
renormalized strength of the Dirac-delta interaction. We choose 
$\lambda_{\cal_R}(\mu^2)$, in the following, in accordance with
the physical value of the pion mass. We observe that,
instead of working  formally with the operator $V^\delta_{\cal R}$, we
could use an ultraviolet momentum cut-off ($\Lambda$), by defining in
this way a regularized interaction. After the construction of the
T-matrix  regularized equation one could perform the limit
$\Lambda \rightarrow \infty $, ariving at the same results as the ones
obtained directly with the use of the renormalized interaction.

The solution of the scattering equation (\ref{trenv}) is found by 
using the two potential formula in terms of the
 T-matrix of the regular potential $V$, $T^V(M^2)$
and the renormalized T-matrix of the Dirac-Delta interaction, see Appendix A, 
which results in
\begin{eqnarray}
&&T_{\cal R}(M^2)=T^V(M^2)
\nonumber \\ 
&+&
\frac{\left( 1+T^V(M^2)G^{(+)}_0(M^2)\right) |\chi>
<\chi| \left(G^{(+)}_0(M^2)T^V(M^2)+1\right)}{
\lambda_{\cal R}^{-1}(\mu^2)- 
<\chi|\left(\frac1{M^2+i\varepsilon-M_0^2 }-
\frac1{\mu^2+i\varepsilon-M^2_0}\right)|\chi>
-< \chi| G^{(+)}_0(M^2)T^V(M^2)G^{(+)}_0(M^2)|\chi>} \ ,
\nonumber \\
\label{trenv2} 
\end{eqnarray} 
where the $T^V(M^2)$ is the solution of the Lippman-Schwinger equation
\begin{eqnarray}
T^V(M^2)=V+VG^{(+)}_0(M^2)T^V(M^2) \ .
\label{tv}
\end{eqnarray}

The structure of Eq.(\ref{trenv2}) allows one more subtraction in the
denominator, which turns faster the convergence of the momentum 
integral in the term where $T^V(M^2)$ is present. This subtraction
is appropriate if the potential $V$ has a Coulomb or Yukawa form,
\begin{eqnarray}
<\vec p| V| \vec q>=\frac{1}{\eta^2 + |\vec p - \vec q|^2} .
\label{trenv3}
\end{eqnarray} 
Thus, we define the renormalized strength of the Dirac-delta interaction
at the subtraction point such that,
\begin{eqnarray}
\lambda_{\cal R}^{-1}(\mu^2)=\overline \lambda_{\cal R}^{-1}(\mu^2)
+< \chi| G^{(+)}_0(\mu^2)T^V(\mu^2)G^{(+)}_0(\mu^2)|\chi> \ ;
\label{trenv4}
\end{eqnarray} 
and introduce the physical information in the renormalized T-matrix
(\ref{trenv2}), through the value of $\overline \lambda_{\cal R}^{-1}(\mu^2)$.

Substituting Eq.(\ref{trenv4}) in (\ref{trenv2}), we obtain
the renormalized T-matrix written as:
\begin{eqnarray}
T_{\cal R}(M^2)&=&T^V(M^2)
\nonumber \\
&+&
\left( 1+T^V(M^2)G^{(+)}_0(M^2)\right) |\chi> t_{\cal R}(M^2)
<\chi| \left(G^{(+)}_0(M^2)T^V(M^2)+1\right)
\ ,
\label{trenv20}
\end{eqnarray} 
where,
\begin{eqnarray}
t^{-1}_{\cal R}(M^2)=\overline \lambda_{\cal R}^{-1}(\mu^2)
- < \chi|\left( G^{V(+)}(M^2)-G^{V(+)}(\mu^2)\right)|\chi> \ ,
\label{trenv21}
\end{eqnarray} 
and the interacting Green's function is
\begin{eqnarray}
G^{V(+)}(M^2)=G^{(+)}_0(M^2)+G^{(+)}_0(M^2)T^V(M^2)G^{(+)}_0(M^2)
\ .
\label{gf}
\end{eqnarray} 

We use the renormalization condition that at the pion mass, $M=m_\pi$, the
T-matrix, for $m_1=m_2$ and $m_1=m_u=m_d$ has the bound-state pole, 
consequently
\begin{eqnarray}
t^{-1}_{\cal R}(m^2_\pi)=0 \ ,
\label{tpi}
\end{eqnarray} 
and  choosing $\mu=m_\pi$ for convenience, which implies that
\begin{eqnarray}
\overline \lambda_{\cal R}^{-1}(m_\pi^2)=0 . 
\label{lpi}
\end{eqnarray} 
The invariance of the renormalized T-matrix (\ref{trenv20}) 
under the dislocation of the subtraction point, just reads that
\begin{eqnarray}
\frac{d}{d\mu^2}t_{\cal R}(M^2)=0 \ ,
\label{cstau}
\end{eqnarray} 
and from (\ref{cstau}) 
\begin{eqnarray}
\overline \lambda_{\cal R}^{-1}(\mu'^2)=
\overline \lambda_{\cal R}^{-1}(\mu^2)
-< \chi|\left( G^{V(+)}({\mu '}^2)-G^{V(+)}(\mu^2)\right)|\chi> \ .
\label{trenv22}
\end{eqnarray} 

At the general subtraction point $t^{-1}_{\cal R}(\mu^2)=
\overline \lambda_{\cal R}^{-1}(\mu^2)$ and  the renormalized
T-matrix at  $\mu^2 $ is given by:
\begin{eqnarray}
T_{\cal R}(\mu^2)&=&T^V(\mu^2)+
\nonumber \\
&&
\left( 1+T^V(\mu^2)G^{(+)}_0(\mu^2)\right) |\chi>
\overline \lambda_{\cal R}(\mu^2)
<\chi| \left(G^{(+)}_0(\mu^2)T^V(\mu^2)+1\right) \ .
\label{trenv5}
\end{eqnarray} 
The full renormalized interaction can be written in a form analogous to
 Eq.(\ref{vfren1}):
\begin{eqnarray}
V_{\cal R}&=&T_{\cal R}(\mu^2)\sum^\infty_{n=0} \left[ -G_0^{(+)}(\mu^2)
T_{\cal R}(\mu^2)\right]^n 
\ , 
\label{trenv6}
\end{eqnarray}
and with that, one could obtain the equation for the renormalized T-matrix
(\ref{trenv}) in the subtracted form (\ref{tvren7}), displayed again below 
\begin{eqnarray} 
T_{\cal R}(M^2)=T_{\cal R}(\mu^2)+T_{\cal R}(\mu^2)
\left( G^{(+)}_0(M^2)- G_0(\mu^2)\right)T_{\cal R}(M^2) \ .
\nonumber 
\end{eqnarray} 
We observe that one could equally well construct the
Callan-Symanzik equation for  $\frac{d}{d\mu^2} T_{\cal R}(\mu^2)$,
by performing the limit of $M\rightarrow\mu$ in (\ref{tvren7}), finding
\begin{eqnarray}
\frac{d}{d \mu^2} T_{\cal R}(\mu^2)=-T_{\cal R}(\mu^2)
\frac{1}{\left(\mu^2+i\varepsilon-M^2_0\right)^2}T_{\cal R}(\mu^2) \ ,
\label{trenv8}
\end{eqnarray}
with the boundary condition given by Eq.(\ref{trenv5}). The solution
of Eq.(\ref{trenv8}) gives the dependence of 
$\overline \lambda_{\cal R}$ on the subtraction point $\mu$ as expressed
by Eq.(\ref{trenv22}).

Now comes a subtle point: It is important to realize that 
the renormalization condition given
by $\overline \lambda_{\cal R}^{-1}(m_\pi^2)=0$, 
and considering Eq.(\ref{trenv21}),
the bare strength of the Dirac-delta interaction is given by
\begin{eqnarray}
   m_r\lambda_{bare}^{-1}=
   \left[< \chi| G^{V(+)}(m_\pi^2)|\chi>\right]_{(m_u,m_{\overline u})} 
.\label{lbare}\end{eqnarray} 
where the Green's function of the Coulomb like interaction
is calculated for equal mass constituent quark and antiquark, i.e.,
$m_1=m_2=m_u=m_{\overline d}$. The bare coupling constant 
in the form explicited by (\ref{lbare}) is sufficient to render
finite the T-matrices from Eq.(\ref{trenv20}).
calculated with different quark masses. 

The bound-state mass, $M_b$, of quark-antiquark states with
also  different quark masses are the position of the
poles of the renormalized T-matrix Eq.(\ref{trenv20}), which
implies
\begin{eqnarray}
m_rt^{-1}_{\cal R}(M_b^2)=
\left[\frac1{m_r}< \chi|G^V(m_\pi^2)|\chi>\right]_{(m_u,m_{\overline u})} 
-\left[\frac1{m_r}< \chi|G^V(M_b^2)|\chi>\right]_{(m_1,m_2)}
=0 \ .
\label{trenv23}
\end{eqnarray} 
The zeroes of $t^{-1}_{\cal R}$ give the position of the
zero angular momentum states of the bound quark-antiquark systems, for
different quark masses and excitation. It is easy to imagine that
in the vicinity of a bound-state of the Coulomb-like potential, because
of the presence  of the pole in $G^V$, the function 
$t^{-1}_{\cal R}$ is rapidly varying and it is infinity at the
Coulomb bound state and because of the change in sign of $G^V$ it
will necessarily presents a zero, and a bound state of the whole
potential.

\section{Pion Charge Radius}

The renormalized effective theory defined by 
Eqs.(\ref{trenv20},\ref{trenv21}) 
with the bound state mass equation (\ref{trenv23}),
once the pion mass is known has only one free interaction parameter, 
$\alpha$, the strength of the effective one-gluon-exchange potential. 
The value of $\alpha$ can be found from the pion charge radius. 
The pion wave-function
in the effective theory comes from the residue of the T-matrix, 
Eq.(\ref{trenv20}), at the pion pole, such that
\begin{eqnarray}
\psi_\pi(x,\vec k_\perp)=\sqrt{\frac{A(k)}{x(1-x)}}\varphi_\pi (\vec k)
\ ,
\label{piwf}
\end{eqnarray}
where
\begin{eqnarray}
\varphi_\pi (\vec k)=<\vec k| G_0(m_\pi^2)
\left( 1+T^V(m^2_\pi)G_0(m_\pi^2)\right) |\chi>
\ ,
\label{piphi}
\end{eqnarray}  
and
\begin{eqnarray}
x(k_z)=\frac12 +\frac{k_z}{2E} \ ,
\label{xkzpi}
\end{eqnarray}
from Eq.(\ref{xkz}).
The absolute normalization of the $q\overline q$ Fock-component of the 
pion wave-function (\ref{piwf}) is such that the asymptotic form is
given by the first term of (\ref{piphi}), with residue at the pion pole
equal one. We impose such normalization condition to be consistent with
the soft pion limit ($m_\pi$=0) for the electromagnetic form factor when 
the Coulomb type interaction goes to zero, as it will be shown below.

\subsection{Including Quark Spin}

The pion electromagnetic form-factor is obtained from the
impulse approximation of the plus component of the current 
$(j^+=j^0+j^3)$ in the Breit-frame with momentum transfer
$q^+=0$ and  $q^2=-{\vec q}^2$. The leptonic decay constant of $\pi^+$
($f_\pi$) is a  physical quantity which depends directly on 
the probability to find the
quark-antiquark Fock-state in the pion wave-function and consequently
properly normalizes it once the empirical value of $f_\pi$ is given. 
Computing the
pion form-factor from an effective Lagrangian, described below, the 
value of $f_\pi$ gives the normalization of the form-factor. In this
case the $q\overline q$ component of the pion wave-function is
normalized  such that, in the vanishing limit of the Coulomb-type 
interaction, it retrieves the asymptotic form Eq.(\ref{softwf})\cite{mill}.
 The coupling of the pion field to the quark field are taken from 
an effective Lagrangian with pseudo-scalar coupling with the
pion quark coupling constant given by the 
Goldberger-Treiman\cite{izuber} relation at the quark level,
\begin{eqnarray}
{\cal L}_{eff}= - i \frac{m}{f_\pi} \vec\pi . 
\overline q \gamma^5 \vec \tau q \ ,
\label{lag}
\end{eqnarray}
 and $f_\pi$= 93 MeV is the pion weak decay constant.

 The general structure of the $q\overline q $ bound state forming the
pion comes from the pseudo-scalar coupling, and we will use such
spin structure in the computation of the Feynman triangle diagram
which expresses the impulse approximation to compute the pion 
electromagnetic current,
\begin{eqnarray}
(p^\mu_\pi+p'^\mu_\pi)F_\pi(q^2)&=&
i 2\frac{m^2}{f^2_\pi} N_c\int \frac{d^4k}{(2\pi)^4}
tr[\frac{\rlap\slash{k}+m}{k^2-m^2+i\varepsilon} \gamma^5 
\nonumber \\ 
&\times&\frac{\rlap\slash{k}-\rlap\slash{p'_\pi}+m}
{(k-p'_\pi)^2-m^2+i\varepsilon} 
\gamma^\mu \frac{\rlap\slash{k}-\rlap\slash{p_\pi}+m}
{(k-p_\pi)^2-m^2+i\varepsilon}  
\gamma^5 ],
\label{triang}
\end{eqnarray}
where $F_\pi(q^2)$ is the pion electromagnetic form-factor, the pion momentum
in the initial and final states are defined by $p^0_\pi=p'^0_\pi$ and 
$\vec {p'}_{\pi\perp}=-\vec p_{\pi\perp}=\vec \frac{q_\perp}{2}$
in the Breit-Frame. $N_c=3$ is the number of colors.

The + component of the current is calculated from (\ref{triang}).
It is chosen because after the integration over $k^-=k^0-k^3$ 
the supression of the pair diagram  is maximal 
for this component in the frame where $q^+=0$  and just the 
valence wave-function enters in the form-factor \cite{mill,pach}.
Although we are going to compute the integration in the - component
of the moment assuming a constant vertex, one can identify in the expression
how the wave-function correspondent to the non-constant vertex 
of Eq.(\ref{piwf}) should be introduced in the expression. The result
is \cite{mill} 
\begin{eqnarray}
F_\pi(q^2)= \frac{2}{(2\pi)^3}\frac{M^2}{f^2_\pi}N_c
\int^1_0dx\int d^2K_\perp 
M^2_0
\left(1+\frac{(1-x)\vec q_\perp.\vec K_\perp}{K^2_\perp+M^2}\right)
\psi_\pi (x,\vec K_\perp)\psi_\pi (x,\vec {K'_\perp}) \ ,
\label{ffactor}
\end{eqnarray}
where the momentum fraction $x=(p^+-k^+)/p^+$. The relative
transverse momentum is given by:
\begin{eqnarray}
\vec K_\perp=(1-x)(\vec p_\perp-\vec k_\perp)-x\vec k_\perp, 
\end{eqnarray}
and $\vec {K'_\perp}= \vec K_\perp +(1-x)\vec q_\perp$.
The free mass operator for the $q-\overline q$ is written in
terms of the momentum fraction and the relative perpendicular 
momentum 
\begin{eqnarray}
 M^2_0=\frac{K^2_\perp+M^2}{x(1-x)}, 
\end{eqnarray}
and $M'_0$ is written as a function of $K'_\perp$.
The expression for the pion form-factor 
gives the standard Drell-Yan formula once the bound-state wave-function 
of the constant vertex model, the asymptotic form, is recognized
\begin{eqnarray}
\psi_\pi(x,\vec K_\perp) = \frac{1}{\sqrt{x(1-x}\left(m^2_\pi-M^2_0\right)},
\label{softwf}
\end{eqnarray}
which is the first term of the pion wave-function in Eq. (\ref{piphi}).
The second term in (\ref{piphi}) comes from the Coulomb-like potential,
and implies in a contribution from the effective theory to the pion
radius. The other factors in Eq.(\ref{ffactor}) comes from 
the Melosh rotations of the individual spin  wave-function of the quarks. 

The pion charge radius from Eq.(\ref{ffactor}) in the soft-pion limit 
 with constant vertex, corresponding to the wave-function of
Eq.(\ref{softwf}) with $m_\pi=0$, gives the well known result of
$r_\pi=\left[ 6 \frac{d}{dq^2}F_\pi(q^2)|_{q^2=0}\right]^\frac12=
\sqrt{3}/(2\pi f_\pi)$ from Ref.\cite{tarr}.  The form factor 
(\ref{ffactor}) in the soft-pion limit with constant vertex, for $q^2=0$, 
reduces to expression for $f_\pi$ \cite{mill}, 
\begin{eqnarray}
f_\pi= \frac{2}{(2\pi)^3}\frac{M^2}{f_\pi}N_c
\int^1_0\frac{dx}{\sqrt{x(1-x)}}\int d^2K_\perp \psi_\pi (x,\vec K_\perp)\ \ ,
\label{fpi}
\end{eqnarray}
obtained from the computation of the leptonic decay transition amplitude 
of $\pi^+$ with the effective Lagrangian (\ref{lag}) and
the wave function given by Eq.(\ref{softwf}).

\subsection{Neglecting Quark Spin}

In order to gain insight on the importance of the inclusion of the quark spin
in the computation of the form-factor, we have simplified 
  the numerator of Eq.(\ref{ffactor}), which is the result from the 
Dirac algebra, by taking the limit of $m\rightarrow \infty$. 
In this way the quark spin
is neglected and the form-factor reduces to the formula found for scalar
particles, and only the overall normalization, which depends on $f_\pi$ is 
mantained. The wave-function is in the form expressed
 by Eq.(\ref{piwf})  with $A(k)=1$. With the above approximations
the pion eletromagnetic form-factor is
\begin{eqnarray}
F_\pi(q^2)= \frac{1}{\pi^3}\frac{M^4}{f^2_\pi}N_c
\int^1_0dx\int d^2K_\perp 
\psi_\pi (x,\vec K_\perp)\psi_\pi (x,\vec {K'_\perp}) \ ,
\label{wsffactor}
\end{eqnarray}
with the relative transverse momentum 
\begin{eqnarray}
\vec K_\perp=(1-x)(\vec p_\perp-\vec k_\perp)-x\vec k_\perp, 
\end{eqnarray}
and $\vec {K'_\perp}= \vec K_\perp +(1-x)\vec q_\perp$.
In case of the  $q\overline q$ wave-function being supossed to be the 
complete pion wave-function, the form factor
is written as \cite{pauli0}
\begin{eqnarray}
F_\pi(q^2)= {\cal N}\int^1_0dx\int d^2K_\perp 
\psi_\pi (x,\vec K_\perp)\psi_\pi (x,\vec {K'_\perp}) \ ,
\label{nwsffactor}
\end{eqnarray}
and from $F_\pi(0)=1$  the normalization $\cal N$ is determined.

The difference between the form factors defined by Eq.(\ref{wsffactor})
and Eq.(\ref{nwsffactor}) are the normalizations of the $q \overline q$
Fock-component of the pion wave-function. In the first case the normalization
is defined by $f_\pi$, while for the second  case
the $q \overline q$ Fock-state component is normalized such that
$F_\pi(0)=1$. As we will see,
Eq.(\ref{nwsffactor}) gives a too small pion radius, while from
Eq.(\ref{wsffactor}) or Eq.(\ref{ffactor}) the pion radius can be described
reasonably. 
The absolute normalization of the form-factor 
computed with the inclusion of the quark spin
 as expressed by Eq.(\ref{ffactor}) 
is undetermined, since the integral diverges. However, the pion radius
is finite with the magnitude of this Fock-component of the 
pion wave-function  known from the empirical value of $f_\pi$, number
of colors and the attributted constituent quark mass.

To close this section, we observe that
our aim is to fit the strength $\alpha$ of the Coulomb-like potential using
the pion charge radius. For this purpose, we introduce 
the pion wave-function from Eq.(\ref{piwf}) in the
form-factor expression, Eq.(\ref{ffactor}) and calculate
the charge radius. Although the form-factor diverges, the 
charge radius is  finite and for $m_\pi=0$
and $\alpha =0$ it retrieves the soft-pion limit. The pion mass and the 
effective Coulomb-like interaction give a correction to the
soft pion limit, $r_\pi^{soft}$=0.58 fm 
towards the experimental result of 0.67$\pm$0.02 fm \cite{amen}.

\section{Numerical Results}

\subsection{Test Case}

We begin the section on the numerical calculations comparing our 
results for the s-wave bound-state energies which are consistent with
 \cite{biel} for the
non-relativistic Coulomb plus repulsive Yukawa model
\begin{eqnarray}
[\epsilon -p^2]\varphi (p)=
\frac1{\pi}\int^\infty_0 dp'\frac{p'}{p}
\left[ \ln\frac{(p-p')^2}{(p+p')^2}-
\ln\frac{(p-p')^2+\eta^2}{(p+p')^2+\eta^2}\right]\varphi (p') \ .
\label{biel}
\end{eqnarray}
In our numerical procedure we checked the results with up
to 200 Gaussian-Legendre quadrature points and the interval
$-1 \ < \ z \ < \ 1$ was transformed to $0 \ < \ k \ < \ \infty$ 
through the variable transformation
$k=c(1-z)/(1+z)$ with $c$ about $1$ .

In Figure 1, we show our results for the first excited state 
$\epsilon^{(2)}$ as a function of the ground state $\epsilon^{(1)}$
and compare with the calculation of \cite{biel} for $\eta=$ 0.1 ,
1 and 10. For these values of $\eta$ we present results in Table I.
Our precision is about 0.5\%, that will suffice for our purposes.
Also the calculation with the renormalized Coulomb plus Dirac-delta
is shown in Figure 1 and Table I. The calculation with non-relativistic
renormalized model corresponds to find the zeroes of the
non-relativistic form of Eq.(\ref{trenv23}),
\begin{eqnarray}
\int d\vec q d\vec p <\vec q|\left( G^{Vnr}(-\epsilon^{(n)}) 
-G^{Vnr}(-\epsilon^{(1)})\right)|\vec p>
=0 \ ,
\label{trenv23nr}
\end{eqnarray} 
with the non-relativistic resolvent $G^{Vnr}(\epsilon)$ for negative
energies obtained from the  solution of 
\begin{eqnarray}
 G^{Vnr}(\epsilon)=G^{nr}_0(\epsilon)+G^{nr}_0(\epsilon)V_CG^{Vnr}(\epsilon)
\ ,
\label{resolnr}
\end{eqnarray} 
where $<\vec p| V_C|\vec q>=\left[\pi^2(|\vec p -\vec q|^2)\right]^{-1}$
and the free resolvent $G^{nr}_0(\epsilon)=[\epsilon-k^2]^{-1}$.

In momentum space Eq.(\ref{trenv23nr}) is given by
\begin{eqnarray}
&&4\pi\int^\infty_0 dp \ p^2 \left[\frac1{-\epsilon^{(n)}-p^2}
-\frac1{-\epsilon^{(1)}-p^2}\right]\nonumber \\
&+&8\pi^2\int^\infty_0 dq \ q^2 \int^\infty_0 dp \ p^2
\left[\frac{t_C(p,q;-\epsilon^{(n)})}{(-\epsilon^{(n)}-p^2)
(-\epsilon^{(n)}-q^2)}
-\frac{t_C(p,q;-\epsilon^{(1)})}{(-\epsilon^{(1)}-p^2)(-\epsilon^{(1)}-q^2)}
\right]
=0 \ ,
\label{p23nr}
\end{eqnarray} 
and s-wave projected T-matrix in Eq.(\ref{p23nr}) is the solution of
\begin{eqnarray}
t_C(p,q;\epsilon)=\frac{1}{2\pi^2}\frac{1}{pq}\ln \frac{(p-q)^2}{(p+q)^2}
+\frac1{\pi}\int^\infty_0dp'\frac{p'}{p}\ln \frac{(p-p')^2}{(p+p')^2}
\frac{t_C(p',q;\epsilon)}{\epsilon-p'^2} \ .
\label{t23nr}
\end{eqnarray}

We solve Eq.(\ref{trenv23nr}) for $n=2$ for each given $\epsilon^{(1)}$.
Now, it is clear the motivation for plotting Figure 1  the
binding energy of the excited state against the ground state energy.
In Table I as well as in Figure 1, the renormalized model reproduce
with less then 10\% accuracy the model results of Eq.(\ref{biel}). In
the limit of $\eta \rightarrow \infty$, the renormalized model
should work better, however in the test case of Eq.(\ref{biel})
this limit corresponds to a vanishing Yukawa interaction, and the
results tends trivially the  Coulomb value of the first excited state
of 0.25 for the ground state value of 1, what in fact  is observed
 in Figure 1 and Table I.

\subsection{Effective Pion Model}

The effective model of \cite{pauli2} corresponds to use
the non-relativistic phase-space $A(k)=1$ in Eq.(\ref{mass1}) and a 
smeared delta-interaction of a Yukawa form:
\begin{eqnarray}
m^2_\pi\varphi (\vec k)=
\left[ 4 m^2+4k^2\right]\varphi(\vec k)-
\frac{4}{3\pi^2}\alpha
\int \frac{d\vec k'}{m}\left(\frac{2m^2}{(\vec k -\vec{k'})^2}+
\frac{\eta^2}{\eta^2+(\vec k -\vec{k'})^2}\right)
\varphi (\vec k') \ ,
\label{mass1c}
\end{eqnarray}
which was solved with parameters adjusted to fit the pion mass and
the rho-meson mass, resulting $m=406$ MeV,
$\alpha =$ 0.6904, and $\eta=$ 1330 MeV \cite{pauli2}. The value
of $\eta$ was found from the condition that the first excited bound state mass
$m^*_\pi$  fullfills the Strutinsk requirement at the extremum
\begin{eqnarray}
\frac{d}{d\eta}m^{*}_\pi=0 \ .
\end{eqnarray}
However, the pion charge radius calculated according to the non-relativistic
formula in \cite{pauli2} is about one half of the empirical value. 
We will confirm this fact, in the renormalized effective model 
while using Eq.(\ref{nwsffactor}) and $\varphi_\pi$ from 
Eq.(\ref{piphi}) with $A(k)=1$ to compute the pion radius.

In Figure 2, we present our results for the pion mass as a
function of $\alpha$ for $\eta=1330$ MeV. Our agreement with the
calculation of \cite{pauli2} is within 10\% . In Figure 3, we
show results for our calculation of the ground and excited
states masses from Eq.(\ref{mass1c}), for
$\alpha$ varying and $\eta=$ 1330 MeV, compared
to the renormalized model for the Coulomb plus Dirac-delta
interaction. In the last case, the bound state masses of 
the pion ground and excited states in s-wave,
are found numerically from the zeroes of Eq.(\ref{trenv23}) with the 
Green's function of the Coulomb-like potential obtained from the
solution of the integral scattering equation (\ref{tv}).
In both equations, which define the non-relativistic 
renormalized model Coulomb plus Dirac-Delta interaction, is used $A(k)=1$. 
We have 
disregarded the effect of the energy transfer in $Q^2$ of Eq.(\ref{q}),
as being a short-range effect parameterized by the value of the
pion mass, which is input in this calculation. 
In momentum space Eq.(\ref{trenv23}) is given by
\begin{eqnarray}
&&4\pi\int^\infty_0 dp \frac{p^2}{A(p)} \left[\frac1{{m^*_\pi}^2-4m^2-4p^2}
-\frac1{m_\pi^2-4m^2-4p^2}
\right]\nonumber \\
&+&8\pi^2\int^\infty_0 dq \frac{q^2}{\sqrt{A(q)}} 
\int^\infty_0 dp \frac{p^2}{\sqrt{A(p)}}
\left[\frac{t^V(p,q;{m^*_\pi}^2)}
{({m^*_\pi}^2-4m^2-4p^2)({m^*_\pi}^2-4m^2-4q^2)} \right.
\nonumber \\ 
&-& \left. \frac{t^V(p,q;m^2_\pi)}{(m^2_\pi-4m^2-4p^2)(m^2_\pi-4m^2-4q^2)}
\right]
=0 \ ;
\label{p23r}
\end{eqnarray} 
and s-wave projected T-matrix in Eq.(\ref{p23r}) is 
\begin{eqnarray}
t^V(p,q;M^2)=\int^1_{-1}dcos(\theta)<\vec p|T^V(M^2)|\vec q> \ ;
\label{t23sw}
\end{eqnarray} 
which is the solution of
\begin{eqnarray}
t^V(p,q;M^2)&=&\frac{4m}{3\pi^2}\frac{\alpha}{pq}
\frac{\ln\frac{(p-q)^2}{(p+q)^2}}{\sqrt{A(p)A(q)}}
\nonumber \\
&+&\frac{8m}{3\pi}\alpha\int^\infty_0\frac{dp'}{\sqrt{A(p)A(p')}}
\frac{p'}{p}\ln\frac{(p-p')^2}{(p+p')^2}
\frac{t^V(p',q;M^2)}{M^2-4m^2-4p^2} \ ,
\label{t23r}
\end{eqnarray} 
the momentum space representation of s-wave projection of Eq.(\ref{tv}).

The agreement between the renormalized model
and the smeared delta model is within a fraction of percent, which 
still improves as the weakly bound limit of the ground state 
is approached, as shown in Figure 3. The calculation of \cite{pauli2} is
few percent below ours. The inclusion of the relativisic
phase-space in the renormalized model of the Coulomb plus
Dirac-delta interaction Eq.(\ref{trenv23}) make less bound
the excited state for a given ground state mass. As the pion mass
grows the relativistic phase-space effect tends to vanish.

In the other study performed,  $\alpha=$ 0.6904 was kept unchanged,
while varying $\eta$ between 350 and 1350 MeV. The results 
for the excited bound state mass against the ground-state mass are
shown in Figure 4. The renormalized Coulomb plus Dirac-delta
calculation agrees within a fraction of percent with the
smeared Dirac-delta calculation. The relativistic
phase-space in the renormalized interaction makes the
excited state less bound for a given ground state mass.

\subsection{Effective Meson Model}

The bound state masses of the meson ground and excited states in s-wave,
are found numerically from the zeroes of Eq.(\ref{trenv23}) with the 
Green's function of the Coulomb-like potential obtained from the
solution of the integral scattering equation (\ref{tv}). The
the energy transfer in $Q^2$ in Eq.(\ref{q}) was neglected. 

The renormalized strength of the singular interaction,
from Eq.(\ref{trenv4}) 
\begin{eqnarray}
\lambda_{\cal R}^{-1}(m^2_\pi)=
< \chi| G_0(m_\pi^2)T^V(m^2_\pi)G_0(m_\pi^2)|\chi> \ ;
\label{lr}
\end{eqnarray} 
with the condition Eq.(\ref{lpi}) at the
physical pion mass is shown in Figure 5 for $m=406$ MeV and
compared to the perturbative calculation
\begin{eqnarray}
\lambda_{\cal R}^{-1}(m^2_\pi)|_{pert}=
< \chi| G_0(m_\pi^2)VG_0(m_\pi^2)|\chi> \ ;
\label{lrpert}
\end{eqnarray} 
with $V$ from Eq.(\ref{mecoul}) and $T^V$ solution of
Eq.(\ref{tv}). For values of $\alpha$ below 0.2 the agreement
between the perturbative and nonperturbative calculation of
the renormalized strength is quite good, giving confidence
to our numerical calculations.

The wave-function $\varphi_\pi$ of the pion from Eq.(\ref{piphi}) 
in momentum space is written as
\begin{eqnarray}
\varphi_\pi (p)=\frac{1}{m^2_\pi-4m^2-4p^2}
\left( \frac{1}{\sqrt{A(p)}}
+2\pi\int^\infty_0 dq \frac{q^2}{\sqrt{A(q)}} \frac{t^V(p,q;m^2_\pi)}
{m^2_\pi-4m^2-4p^2}
\right) 
\ .
\label{piphip}
\end{eqnarray}  
The first-order perturbative pion wave-function is
\begin{eqnarray}
\varphi_\pi (p)|_{pert}=\frac{1}{\sqrt{A(p)}}\frac{1}{m^2_\pi-4m^2-4p^2}
\left( 1+
\frac{8m}{3\pi}\alpha\int^\infty_0 \frac{dq}{A(q)}\frac{q}{p} 
\frac{\ln \frac{(p-q)^2}{(p+q)^2}}{m^2_\pi-4m^2-4p^2}
\right) 
\ ,
\label{piphipert}
\end{eqnarray}  
and for $\alpha=0$ the pion wave-function has
the asymptotic form
\begin{eqnarray}
\varphi^0_\pi (\vec p)=\frac{1}{\sqrt{A(p)}}\frac{1}{m^2_\pi-4m^2-4p^2}
\ .
\label{piphi0}
\end{eqnarray}  

In \cite{pauli2}, the pion wave-function eigenfunction of the effective 
square mass operator of the Coulomb plus Yukawa
model, Eq.(\ref{mass1c}), was approximated by the following analytical
form:
\begin{eqnarray}
\varphi^{a}_\pi (\vec p)=\frac{\cal N}{\left( 1+p^2/p_a^2\right)^2}
\ ,
\label{piphia}
\end{eqnarray}  
where $\cal N$ is an arbitrary normalization and the fit is
performed for $\alpha=0.6904$, $\eta=$ 1330 MeV and $m$=406 MeV
and  $p_a$=515 MeV is adjusted to the numerical solution of Eq.(\ref{mass1c}).

The various non-relativistic models of the pion wave-function, with $A(k)$=1,
 are plotted in Figure 6. In this figure, the comparision 
between $\varphi_\pi$, $\varphi_\pi |_{pert}$ , 
$\varphi^a_\pi$ and $\varphi^0_\pi$ are performed. The model parameters are 
$\alpha=0.6904$ and $m=$406 MeV and the normalization is arbitrary. 
The perturbative calculation reproduce $\varphi_\pi (p)$ for $p$ below $ m$. 
The asymptotic wave-function $\varphi^0_\pi (p)$  overestimate
$\varphi_\pi (p)$, as it should be for a repulsive Coulomb-type
interaction, and  for small $p$ it approaches the nonperturbative 
eigenfunction. The analytical approximation works quite good for 
momentum up to about 1.5$m$. More results on the pion wave-function
are shown in Figure 7, where the results for $A(k)=1$ with the full 
calculation are compared. The effect of $A(p)$ diminuishes the magnitude of
 $\varphi(p)$, as one could antecipate from Eq.(\ref{piphip}). Also
we plot $\varphi_\pi(p)$ for $\alpha= \ 0.18$ and $m= \ 386$MeV, which
is above the curve of previous case due to the decrease of the
Coulomb attraction (this parameters fits the empirical pion radius).
We point also that, for very high momentum, again the asymptotic form
dominates, as the second term of Eq.(\ref{piphip}) tends to zero faster
than the first one.

We are going to compute the pion radius in the effective renormalized model
of the pion using Eq.(\ref{ffactor}),
Eq.(\ref{wsffactor}) and Eq.(\ref{nwsffactor}). In the
last two cases, the pion wave-function from (\ref{piphip}) with
$A(k)=1$ is used.
The pion wave-function is known from Eqs.(\ref{piwf}) and 
(\ref{piphip}) from which the charge radius including the effect of quark spin
is obtained from integration by Gaussian quadrature of the difference 
\begin{eqnarray}
\Delta F_\pi(q^2)=F_\pi (q^2)-F_\pi (0) \ ,
\end{eqnarray}
obtained from Eq.(\ref{ffactor}),
and the pion radius is calculated from
\begin{eqnarray}
   r_\pi=\left[6\frac{d}{dq^2}\Delta F_\pi(q^2)|_{q^2=0}\right]^\frac12
\ .\end{eqnarray}
In the case the quark spin is neglected, two possibilities 
of calculation of the pion radius are considered. One through 
Eq.(\ref{wsffactor}) in which only the spin factors are simplified
in the limit of the quark mass being infinity, while the normalization
is defined as in Eq.(\ref{ffactor}), known from the empirical value 
of $f_\pi$ and number of colors. The second possibility is to
declare normalized to one the $q\overline q$ Fock-component of
the pion wave-function and use the formula for the form-factor
in which the quark has no spin, Eq.(\ref{nwsffactor}).

The results for the pion charge radius as a function of the
strength $\alpha$ of the Coulomb-like interaction, Eq. (\ref{mecoul}),
are shown in Figure 8. For $r_\pi=0.67$fm we found $\alpha$=0.18
using Eq.(\ref{ffactor}).
Our calculations obtains the pion charge radius
from the relativistic expression (\ref{ffactor}) which is known
to give about twice the non-relativistic radius \cite{mill}. 
The attractive Coulomb like interaction increases the radius over
the soft-pion limit with $\alpha=0$ which is below its experimental value.
In that sense a consistence is found with the effective theory which
has an attractive Coulomb-like interaction. The repulsion would be completely
inconsistent with  the pion radius.

The pion charge radius obtained from first order  perturbative
calculation of the pion wave-function
\begin{eqnarray}
\psi_\pi(x,\vec k_\perp)|_{pert}=\sqrt{\frac{A(k)}{x(1-x)}}
\varphi_\pi (k)|_{pert}
\ ,
\label{piwfpert}
\end{eqnarray}
is also shown in Figure 8. In consistency with the strength calculation
presented in Figure 5, we observe that
for $\alpha$ below 0.2 the perturbative calculations  
match the nonperturbative results.

The results for the scalar quarks form-factors with $\varphi$ from
Eq.(\ref{piphip}) calculated in the effective renormalized theory
for $A(k)=1$ are also shown in Figure 8. The calculation of $r_\pi$
with Eq.(\ref{wsffactor}) show values above the ones calculated with
Eq.(\ref{nwsffactor}), which are two small  compared to the empirical 
value, indicating the importance of the physical normalization of
the $q\overline q$ Fock-component of the wave-function using $f_\pi$.
We also performed the first order perturbative calculation of $r_\pi$
for the wave-function normalized  to 1. The agreement between the 
perturbative nonperturbative calculation is reasonable for $\alpha$ below
0.2 .

The plot of the pion charge radius against the 
mass of the first excited-state in shown in Figure 9.
For decreasing values  of $\alpha$, the pion charge radius
diminuishes and consistently the excited-state mass increases,
i.e., this state becomes less bound. It is clear from this figure
that to simultaneously fit the radius and the mass of the rho-meson
(768 MeV) we are obliged to use a different quark mass from the
value of 406 MeV. For comparision we also show the results from
Eq.(\ref{wsffactor}).

The experimental pion radius of 0.67$\pm$0.02 fm is fitted with
$m_u=m_{\overline u}=$ 386 MeV and $\alpha=$ 0.18 resulting
$r_\pi=$ 0.67 fm and the mass of the singlet-2s state of 768 MeV,
we remind that the pion mass is input in the renormalized model
calculation. The singlet-2s excited state mass of the 
$\overline u u$ system is
identified with the $\rho$ meson mass in the present effective QCD
model. The quark mass is varied to form  mesons with 
one up antiquark together with the strange, 
charm or bottom quarks. The masses of the constituent quarks 
were within the range of 500 to 5000 MeV. The results are shown in 
Table II and compared to the experimental data. The singlet-2s states are 
identified with the lowest mass vector mesons states, since the
Dirac-delta interaction is the effective hyperfine interaction,
the reason for the spliting between the pseudo-scalar and vector mesons.
Although, in the singlet channel the hyperfine interaction is
attractive, which is not valid for the spin one mesons, 
we believe that the Dirac-delta interaction mock up 
short-range physics beyond that, by taking care of the empirical value 
of the pion mass.

It is clear that the split between
the heavy meson masses will be not adjusted in the present calculation
for $\alpha$ small.
This is reasonable, since  these mesons are weakly bound and the 
wave-function spreads out in the region where the confinement potential,
not present in our model, should be important. In
Figure 10, we study the difference of the excited and ground state masses
as a function of the ground state mass. As we have seen in Table II,
the difference is underestimated for $\alpha=$ 0.18 above the kaon mass.
For $\alpha=0.5$ the experimental data is reproduced. This
is in fact reasonable if we think that  $\alpha$ should on average
increase with the size scale, indicating the confininig behaviour,
which is stronger for the heavy mesons
since they are less bound than the pion and the kaon. 

\section{Conclusion}

The essential development made in this work is the
renormalization of the effective QCD-inspired Hamiltonian theory with 
a singularity at zero range and its consequent application to the
pion and other mesons. The method is 
an example of the Hamiltonian renormalization procedure and it
is equivalent to a subtracted equation for the transition matrix.
The physical renormalization condition is given to the two-particle model at 
the subtraction point and in the cases discussed here it is 
the ground state binding energy or mass.
The treatment is shown to be renormalization group invariant, i.e.,
independent on the arbitrary subtraction point. This independence is expressed
by a fixed-point Hamiltonian that brings the physical input to the theory, 
the pion mass or ground state binding energy, as well as the 
necessary counterterms that render all the momentum integrations finite. 

First, we have studied in an example, 
the renormalization method applied to a two-body model with
a Coulomb plus Dirac-delta interaction, where we have calculated 
the excited state energy for a given ground state binding energy. The results
are compared to calculations with a Coulomb plus repulsive Yukawa interaction 
and the renormalized model reproduce with less then 10\% accuracy the 
model results of Eq.(\ref{biel}). The success of this result 
drive us to the solution of the renormalization problem of the 
$\uparrow\downarrow$-model\cite{pauli2}.
  
The effective mass operator equation in the
$\uparrow\downarrow$-model Eq.(\ref{p1}), has as eigenstate 
the lowest Fock-state 
component ($\overline q q$) of the light-cone wave-function of a 
meson bound system of constituent quarks or dressed quarks (not to be confused 
with the bare quark). This model picks out one particular aspect of the gluon
exchange between quarks, 
namely the strong attraction of the spin-spin interaction in
the singlet channel. Previously \cite{pauli2}, the renormalization has 
been carried by first regularizing the Dirac-delta interaction through 
a Yukawa form and then its parameter was found by the Strutinsky 
requirement that the mass of the excited state is stationary in respect 
to variation of the regularization parameter. The step forward in this
work was the use of the renormalization group invariant approach, in which
the regularization parameter is not necessary to solve the model. All
the short-range physics is parametrized by one parameter: the renormalized
strength of the Dirac-delta interaction, which is determined by the mass
of the pion. We showed that,  the results for the mass of the excited 
state obtained with the renormalized model and the smeared delta 
regularized model are in agreement within  fraction of percent,
 for the same ground state mass,  
which was varied either by changing the Coulomb interaction intensity or
the Yukawa range. The concordance still improves as the weakly 
bound limit of the ground state is approached.  The effect of 
the relativisic phase-space in the renormalized model of the Coulomb plus
Dirac-delta interaction Eq.(\ref{trenv23}) makes the excited state, 
for a given ground state mass, less bound. This difference tends to vanish
as the pion becomes weakly bound. 

The various models of the $\overline qq$ Fock-component 
of the pion light-cone wave-function, considering $A(k)$=1
and the relativistic phase-space ($A(k)$ from Eq.(\ref{phsp})), were
calculated and compared with the analytical form and perturbative
result. We obtained the reduced wave-function
$\varphi_\pi$ (\ref{piphip}) solution of the renormalized 
$\uparrow\downarrow$-model,
 $\varphi_\pi |_{pert}$ (\ref{piphipert}), the 
analytical form $\varphi^a_\pi$ (\ref{piphia}) and the asymptotic form 
$\varphi^0_\pi$ (\ref{piphi0}). We have used  model parameters 
$\alpha=0.6904$ and $m=$406 MeV \cite{pauli2}. Although for
for $p$ below $ m$ all the calculations are reasonably consistent, the
high momentum tail is dominated by the asymptotic wave-function 
$\varphi^0_\pi (p)$, which overestimate
$\varphi_\pi (p)$, as it should be for a repulsive Coulomb-type
interaction, and for small $p$ it approaches the nonperturbative 
eigenfunction, as well as for very high momentum. The analytical 
approximation works quite good for momentum up to about 1.5$m$,
however it does not have the asymptotic tail for high momentum. 
The relativistic phase-space  diminuishes the 
magnitude of $\varphi(p)$ as clearly seen in Eq.(\ref{piphip}). 

The calculation of the pion charge radius was performed in
the renormalized effective QCD-inspired Hamiltonian theory
with Eqs. (\ref{piphi}) and (\ref{ffactor})
and although we have simplified the spin dependence in
the dynamical equation it is important in the evaluation of the radius.
For this purpose we have used an effective pseudo-scalar
Lagrangian to  construct the spin part of the pion wave-function, 
it gives the absolute normalization of the lowest Fock-component of the
light-cone wave-function in terms of the weak decay constant $f_\pi$,
 the constituent quark mass and number of colors. Turning off the 
Coulomb-like interaction, we retrieved the well
knwon result valid in the soft pion limit\cite{tarr}. We also compared
the values for $r_\pi$ obtained
with the expressions where the quark spin is neglected and the wave-function
normalized to 1. We  pointed out the necessity of the  
correct normalization of the wave-function 
of the $\uparrow\downarrow$-model according to the $f_\pi$ value to
be able to fit $r_\pi$,
which is reminiscent of the fact that $f_\pi$ and $r_\pi$ are closely 
related in the light-front phenomenology \cite{mill}. 

The experimental pion radius of 0.67$\pm$0.02 fm was fitted with
$m_u=m_{\overline u}=$ 386 MeV and $\alpha=$ 0.18 resulting
$r_\pi=$ 0.67 fm and the mass of the singlet-2s state of 768 MeV,
we remind that the pion mass is input in the renormalized model
calculation. The singlet-2s excited state mass of the 
$\overline u u$ system was identified with the $\rho$ meson 
mass in the present effective QCD model. We 
stress that in the singlet channel the hyperfine interaction is
attractive, which is not valid for the spin one channel of the vector mesons,
 however, we believe that the Dirac-delta interaction mock up 
short-range physics beyond that, which is brought to the model
by  the empirical value of the pion mass. In essence, without being too
naive, in order to fit the pion charge radius
with sucess and with reasonable parameters, it was essential that
 {\it i)} the Coulomb-like interaction be {\it attractive}
 and {\it ii)} the {\it normalization} 
of the form-factor be consistent with the empirical $f_\pi$ value.

In respect to the masses of the ground state of the
pseudo-scalar and vector mesons with 
one up antiquark together with the strange, 
charm or bottom quarks, which were calculated 
as a function of the constituent quark mass, the results were 
in qualitative agreement with the data for  $\alpha=0.18$. 
We tried a better fit of this data, since it was clear that the 
split between the heavy meson masses would not be adjusted 
in the present calculation for $\alpha$ small. This in fact, seems
reasonable, since  these mesons are weakly bound and the 
quarks can be found in a region where the confinement potential,
not present in our model, is important.
The difference between the vector and pseudo-scalar mesons 
masses for the same $\overline q q$ pair is underestimated for $\alpha=$ 0.18 
above the kaon mass. We found that for $\alpha=0.5$ the experimental 
data is reproduced. This gives us some hope that it is possible to 
refine the $\uparrow\downarrow$-model to include more physics 
than initially thought. In  regard to this extension, it is
reasonable to think that  $\alpha$ should on average
increase with the size scale, indicating the confininig behaviour,
which we found stronger for the heavy mesons
since they are less bound than the pion and the kaon. In short,
a reasonable description of the physics of the pion and other
scalar and vector mesons were found,
taking into account the simplicity of the renormalized
effective Light-Cone QCD-inspired theory.

{\bf Acknowledgments:} 
We thank ECT$^*$ for the kind hospitality during the
"International Workshop on Relativistic Dynamics and Few-Hadron Systems" 
from 6 to 17 Nov 2000 where this work has been initiated. 
TF also thanks CNPq and FAPESP for financial support.

\newpage
\appendix

\section{Derivation of the Renormalized T-Matrix}

In this Appendix, the solution of  Eq.(\ref{trenv}) to find the
the renormalized T-matrix, Eq.(\ref{trenv2}), is performed in detail.
We want to solve Eq.(\ref{trenv})
\begin{eqnarray}
T_{\cal R}(M^2)=V+ V_{\cal R}^\delta 
+\left(V+ V_{\cal R}^\delta \right) G^{(+)}_0(M^2)T_{\cal R}(M^2)) \ .
\label{atrenv}
\end{eqnarray} 
The regular part of the potential, $V$, is defined by Eq.(\ref{mecoul})
and the  renormalized singular interaction is given by Eq.(\ref{vfren1}),
rewritten as a matrix equation 
\begin{eqnarray}
V_{\cal R}^\delta =|\chi> \lambda_{\cal R}(\mu^2)<\chi|
-|\chi> \lambda_{\cal R}(\mu^2)<\chi|G^+_0(\mu^2)V_{\cal R}^\delta \ ;
\label{avd}
\end{eqnarray}
which has the solution
\begin{eqnarray}
V_{\cal R}^\delta =|\chi> v_{\cal R}(\mu^2)<\chi| \ ,
\label{avd1}
\end{eqnarray}
with the function
\begin{eqnarray}
v_{\cal R}(\mu^2)=\left[\lambda_{\cal R}(\mu^2)+<\chi|G^+_0(\mu^2)|\chi>
 \right]^{-1}\ .
\label{avv}
\end{eqnarray}
The function $v_{\cal R}(\mu^2)$ contains the divergences in the
momentum integrals which exactly cancels  such  infinities in 
Eq.(\ref{atrenv}). It is enough for the formal manipulations
that will come. However, one could equally well introduces a cutoff
in Eqs. (\ref{atrenv}) and (\ref{avv}), and performs the limit of the
cutoff going to infinity just after the solution of Eq.(\ref{atrenv}),
in which all the necessary cancellations happens and the limit
is finite.

Next, Eq.(\ref{atrenv}) is rewritten as:
\begin{eqnarray}
\left(1-VG^{(+)}_0(M^2)\right)T_{\cal R}(M^2)=V+ V_{\cal R}^\delta 
+V_{\cal R}^\delta G^{(+)}_0(M^2)T_{\cal R}(M^2)) \ ,
\label{atrenv1}
\end{eqnarray} 
and inverting the operator in the left-hand 
using the regular T-matrix, $T^V(M^2)$, solution of Eq.(\ref{tv}),
one has
\begin{eqnarray}
T_{\cal R}(M^2)&=&T^V(M^2) \nonumber \\ 
&+&\left( 1+T^V(M^2)G^{(+)}_0(M^2)\right)|\chi>v_{\cal R}(\mu^2)
<\chi|\left(G^{(+)}_0(M^2)T_{\cal R}(M^2)+1\right) \ .
\label{atrenv2}
\end{eqnarray} 

The "bra" function, $<\chi|G^{(+)}_0(M^2)T_{\cal R}(M^2)$, has to be calculated
in order to find the renormalized T-matrix. We multiply 
 Eq.(\ref{atrenv2}) by $<\chi|G^{(+)}_0(M^2)$ on both sides, and solving it
we get
\begin{eqnarray}
<\chi|G^{(+)}_0(M^2)T_{\cal R}(M^2)=
\frac{<\chi|G^{(+)}_0(M^2)T^V(M^2)+
<\chi|G^{V(+}(M^2)|\chi>v_{\cal R}(\mu^2)<\chi|}
{1-<\chi|G^{V(+)}(M^2)|\chi>v_{\cal R}(\mu^2)} \ ,
\label{bra}
\end{eqnarray}
where regular potential resolvent is
\begin{eqnarray}
G^{V(+)}(M^2)=G^{(+)}_0(M^2)+G^{(+)}_0(M^2)T^V(M^2)G^{(+)}_0(M^2) \ .
\label{agv}
\end{eqnarray}

The "bra" function of Eq.(\ref{bra}) is introduced back in 
Eq.(\ref{atrenv2}), and with a little algebra one finds 
\begin{eqnarray}
&&T_{\cal R}(M^2)=T^V(M^2)
+
\frac{\left( 1+T^V(M^2)G^{(+)}_0(M^2)\right) |\chi>
<\chi| \left(G^{(+)}_0(M^2)T^V(M^2)+1\right)}
{v_{\cal R}^{-1}(\mu^2)-<\chi|G^{V(+)}(M^2)|\chi>} \ ,
\label{atrenv3} 
\end{eqnarray} 
which, after introducing Eq.(\ref{avv}) and the explicit form
of the resolvent of Eq.(\ref{agv}), results in 
the renormalized T-matrix of Eq.(\ref{trenv2}).

\newpage

\begin{table}
\begin{tabular}{cccc} 
%\hline
$\eta$ & $ \epsilon^{(1)}$ [C-Y] & $\epsilon^{(2)}$ [C-Y] 
& $\epsilon^{(2)}$  [C-$\delta$] 
\\ \hline 
0.1 & 0.1109 & 0.06781 & 0.06237  \\
1   & 0.5119 & 0.1813  & 0.1736   \\
10  & 0.9495 & 0.2449  & 0.2439 \\
\hline 
\end{tabular} 
\caption{ Results for S-wave binding energies of the 
non-relativistic model of Eq.(\protect{\ref{biel}}), 
our calculation with $N=100$ [C-Y], compared to 
the effective Coulomb plus Dirac-delta interaction  [C-$\delta $]. }
\vspace{2 cm}
\begin{tabular}{ccccccc} 
%\hline
$\overline q  q$ &$m_1$ & $m_2$ & $M_{b,th}^{(1s)}$   & 
$M_{b,exp}^{(1s)}$ &$M_{b,th}^{(2s)}$  &$M_{b,exp}^{(2s)}$ 
\\ \hline 
$\overline ud$& 386 & 386  & 140   & $\pi^\pm : $ 140 & 768 & $\rho : $ 768 \\
$\overline u s $& 386 & 500 & 511  & $ K^\pm : $   494 & 882 &$ K^*: $ 892 \\
$\overline u c$& 386 & 1500 & 1852  & $D^0 :$ 1865& 1882& $D^{*0}:$ 2007 \\
$\overline u b$& 386 & 5000 & 5375  & $B^\pm:$ 5279& 5383& $B^*:$ 5325  \\
\hline 
\end{tabular} 
\caption{Results for S-wave Meson Masses: singlet-1s $(M^{(1s)}_{b,th})$ and 
singlet-2s $(M^{(2s)}_{b,th})$ and experimental values $(M^{(1s,2s)}_{b,exp})$.
All masses in MeV. $\alpha$ =0.18.}
\end{table}

%%%%%%%%%%%%%%%%%%%%%%%%%%%%%% FIGURES AND CAPTIONS %%%%%%%%%%%%%%%%%%%%%%%%%% 
%%%%%%%%%%%%%%%%%%%%%%%%%%%%%% FIG. 1 %%%%%%%%%%%%%%%%%%%%%%%%%%%%%%%%%%%%%%%% 
\vskip -2cm 
{\small{ 
\begin{description} 
\item[Fig. 1]  
\postscript{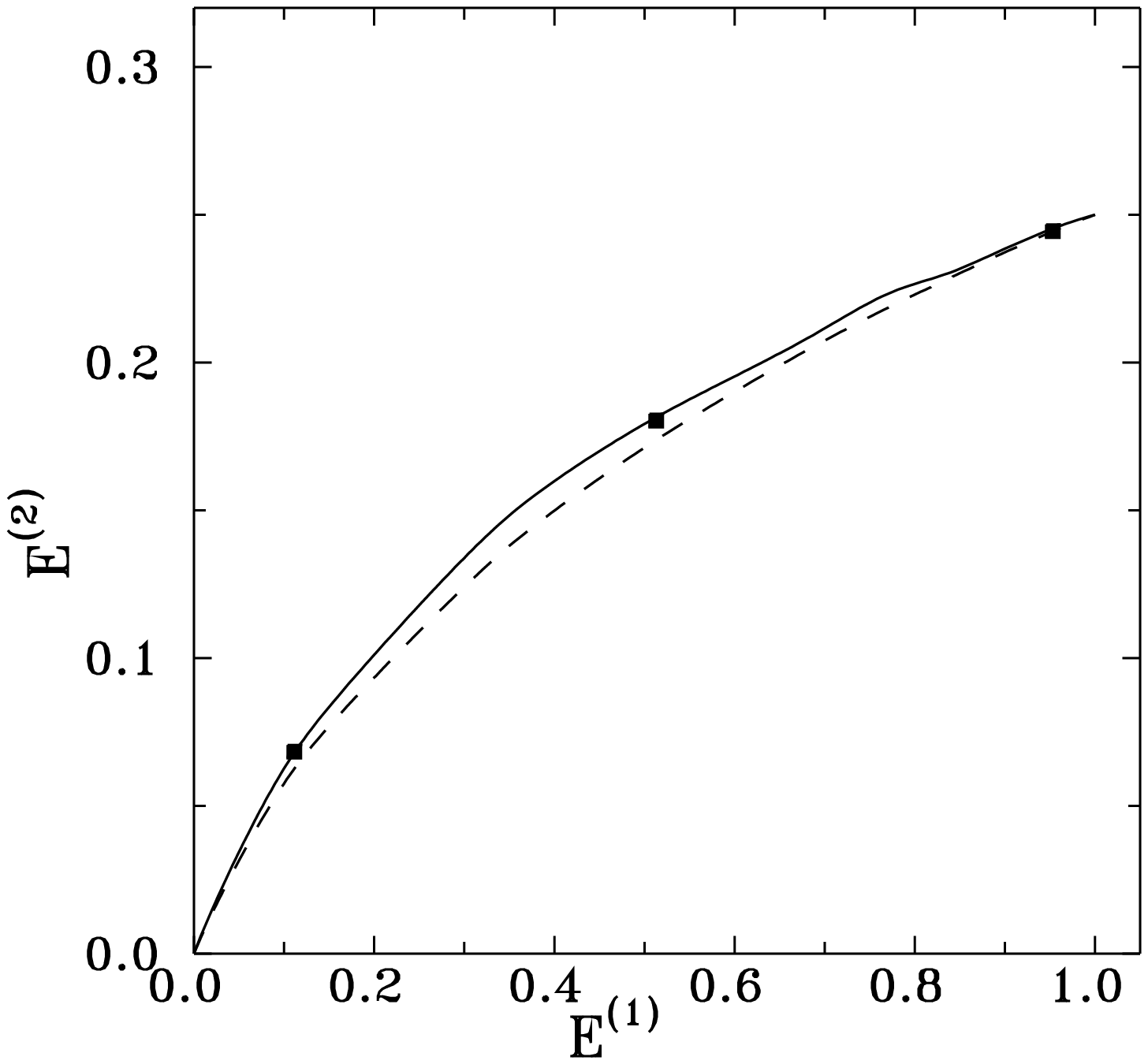}{0.8}    %**epsf** 
 \vskip -1cm 
Excited state binding energy ($\epsilon^{(2)}$) as a function of the ground 
state binding energy ($\epsilon^{(1)}$) for the non-relativistic model.
Attractive Coulomb plus repulsive Yukawa 
non-relativistic model of the Ref. \cite{biel} (full square). Our numerical
calculation of Eq.(\ref{biel})
is given by the solid line, and the solution of the 
effective model Eq.(\ref{trenv23nr}) (Coulomb plus Dirac-Delta) is given 
by the dashed line.
%%%%%%%%%%%%%%%%%%%%%%%%%%%%%% FIG. 2 %%%%%%%%%%%%%%%%%%%%%%%%%%%%%%%%%%%%%% 
\vspace{1 true cm} 
\item[Fig. 2]  
\postscript{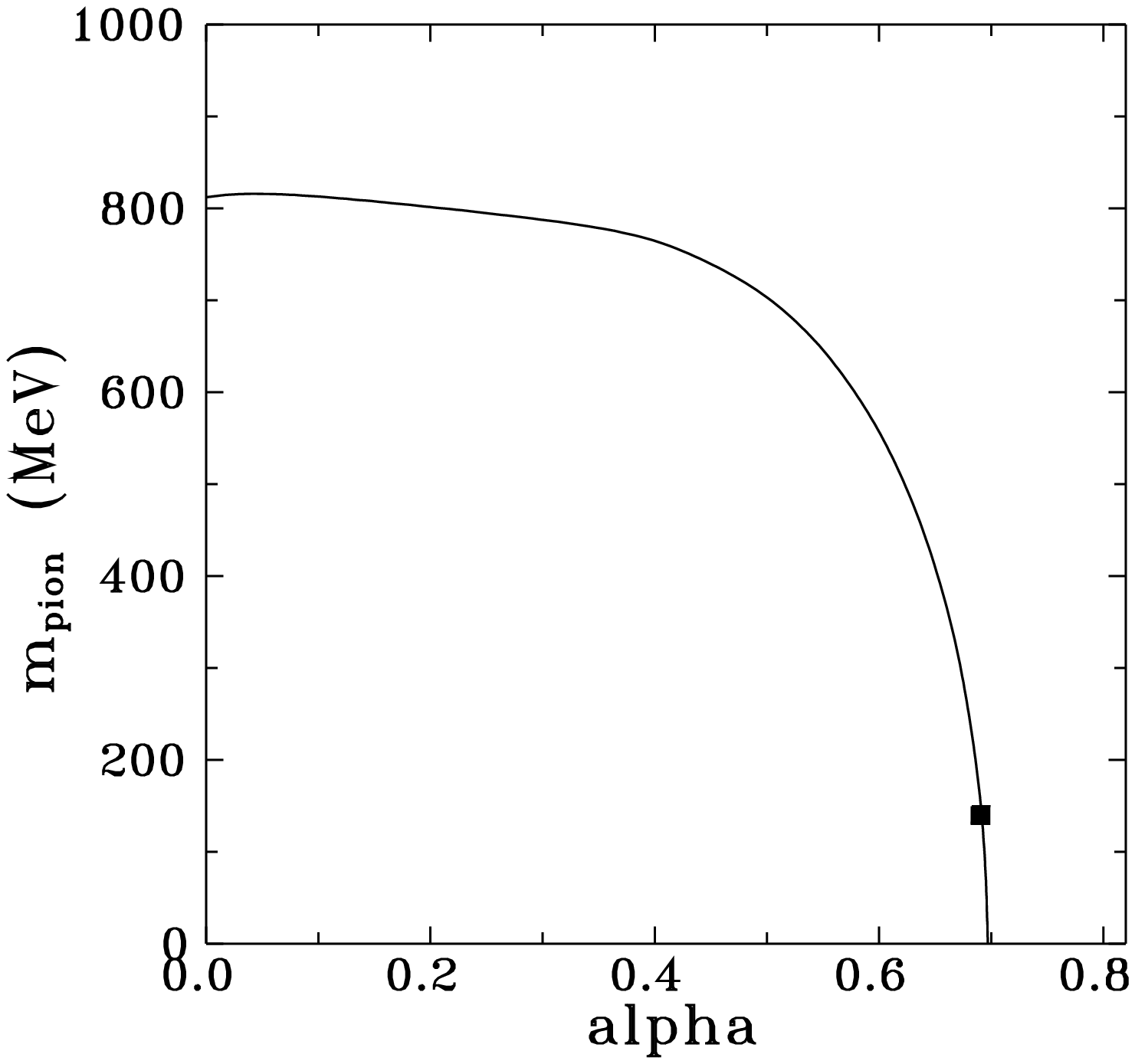}{0.8}    %**epsf** 
 \vskip -1cm 
Pion mass as a function of the strength of the Coulomb interaction $\alpha$.
Non-relativistic model for Coulomb plus Yukawa interaction, $m=$ 406 MeV, 
solid line. Yukawa range parameter $\eta$=1330 MeV\cite{pauli2}. 
The full square is the calculation of Ref. \cite{pauli2}.
%%%%%%%%%%%%%%%%%%%%%%%%%%%%%% FIG. 3 %%%%%%%%%%%%%%%%%%%%%%%%%%%%%%%%%%%%%% 
\vspace{1 true cm} 
\item[Fig. 3]  
\postscript{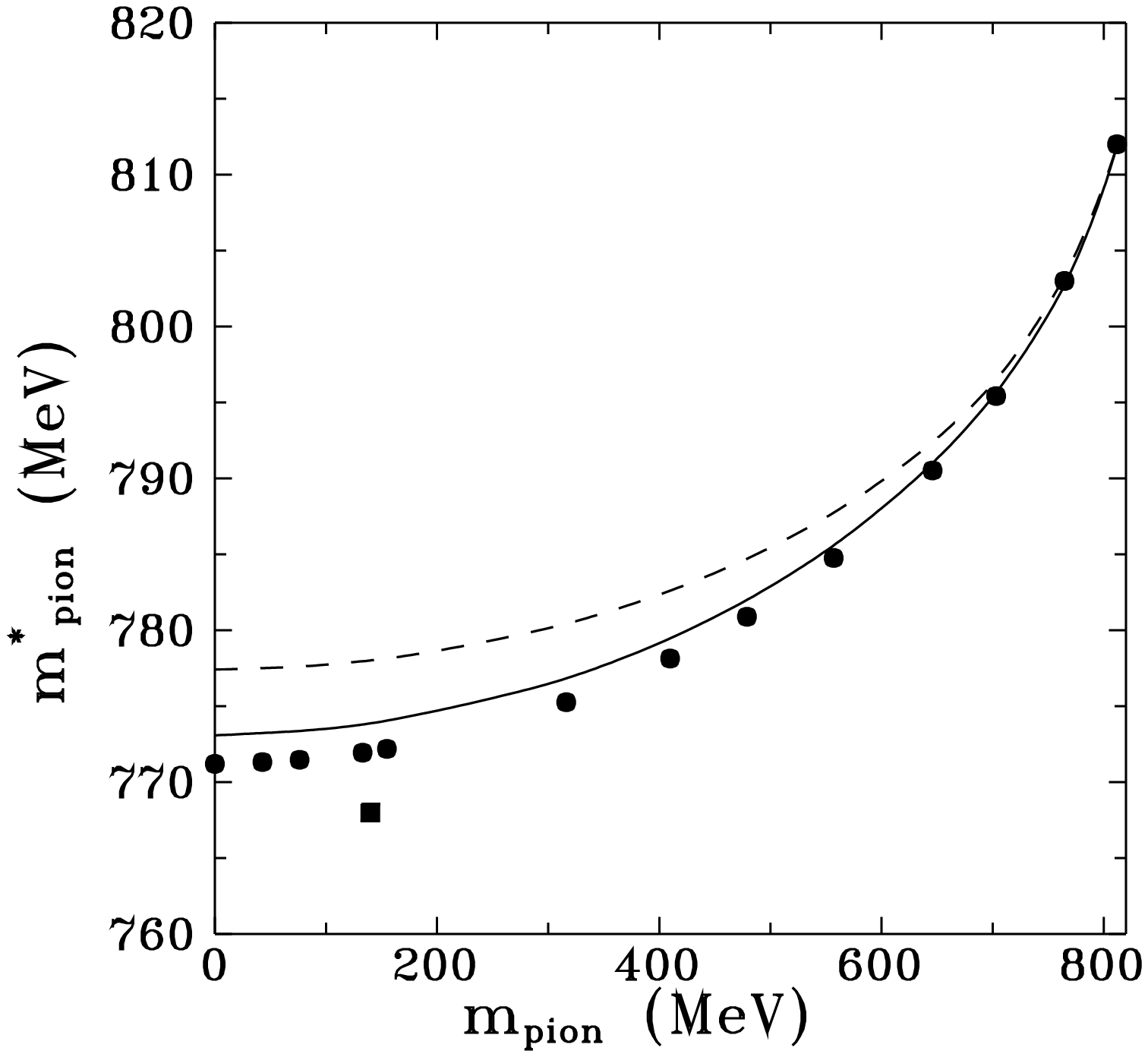}{0.8}    %**epsf** 
  \vskip -1cm 
Mass of the excited state as a function of the ground state pion mass.
Non-relativistic model for Coulomb plus Yukawa interaction, $m=$ 406 MeV, 
solid line. Strength of the Coulomb potential within $0 <\alpha <0.7$ and
Yukawa range parameter $\eta$=1330 MeV\cite{pauli2}. 
The full square is the calculation of Ref. \cite{pauli2}.
The non-relativistic effective model (Coulomb plus Dirac-Delta)
is given by the full dots. 
The relativistic effective model for the Coulomb plus Dirac-Delta interaction
is given by the dashed-line. 
%%%%%%%%%%%%%%%%%%%%%%%%%%%%%% FIG. 4 %%%%%%%%%%%%%%%%%%%%%%%%%%%%%%%%%%%%%% 
\vspace{1 true cm} 
\item[Fig. 4]  
\postscript{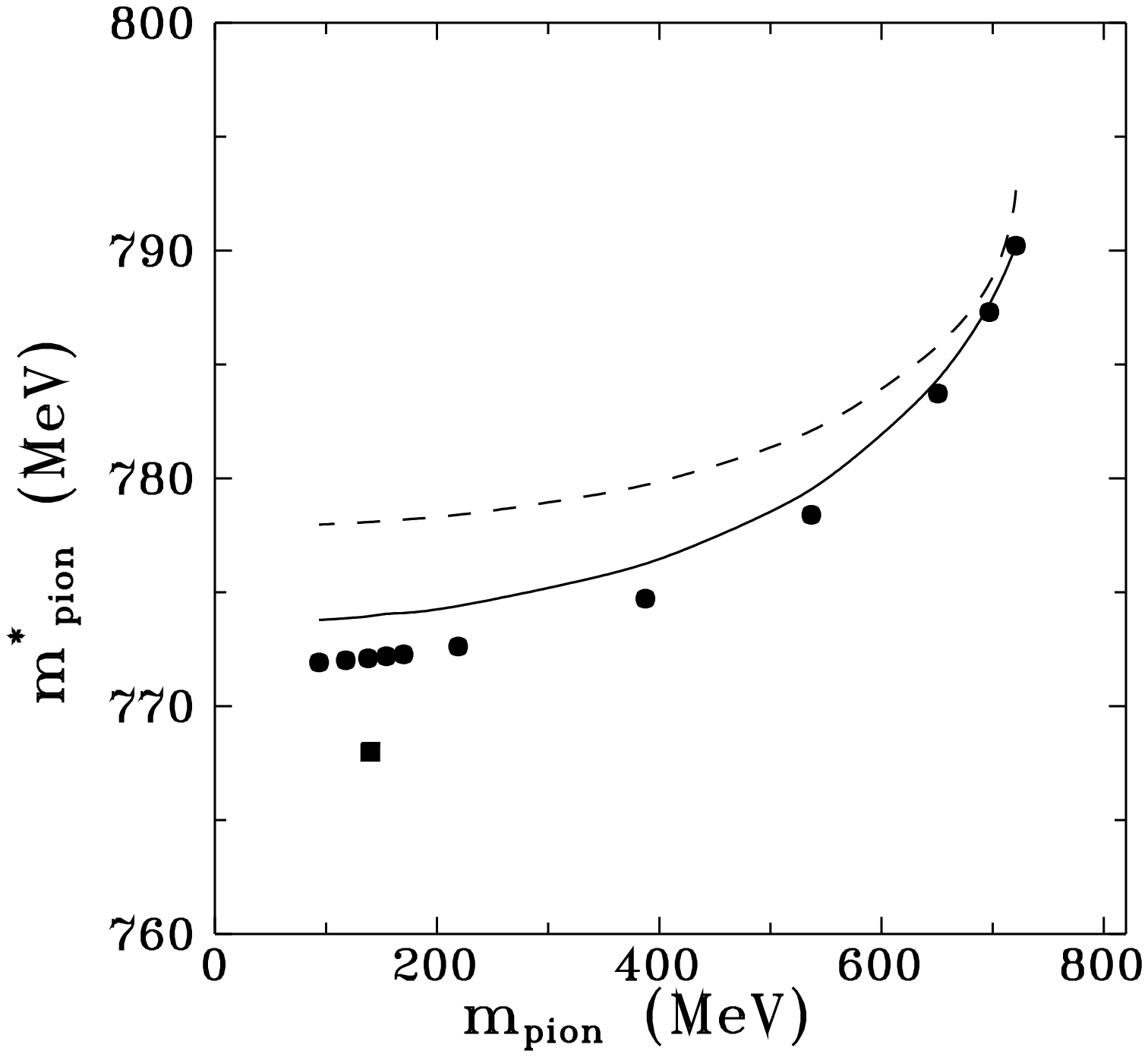}{0.8}    %**epsf** 
  \vskip -1cm 
Mass of the excited state as a function of the ground state pion mass.
Non-relativistic model for Coulomb plus Yukawa interaction, $m=$ 406 MeV, 
solid line. Strength of the Coulomb potential $\alpha = 0.6904$ and
Yukawa range parameter within $0<\eta<1350$ MeV \cite{pauli2}. 
The full square is the calculation of Ref. \cite{pauli2}.
The non-relativistic effective model (Coulomb plus Dirac-Delta)
is given by the full dots. 
The relativistic effective model for the Coulomb plus Dirac-Delta interaction
is given by the dashed line. 
%%%%%%%%%%%%%%%%%%%%%%%%%%%%%% FIG. 5 %%%%%%%%%%%%%%%%%%%%%%%%%%%%%%%%%%%%%% 
\vspace{1 true cm} 
\item[Fig. 5]  
\postscript{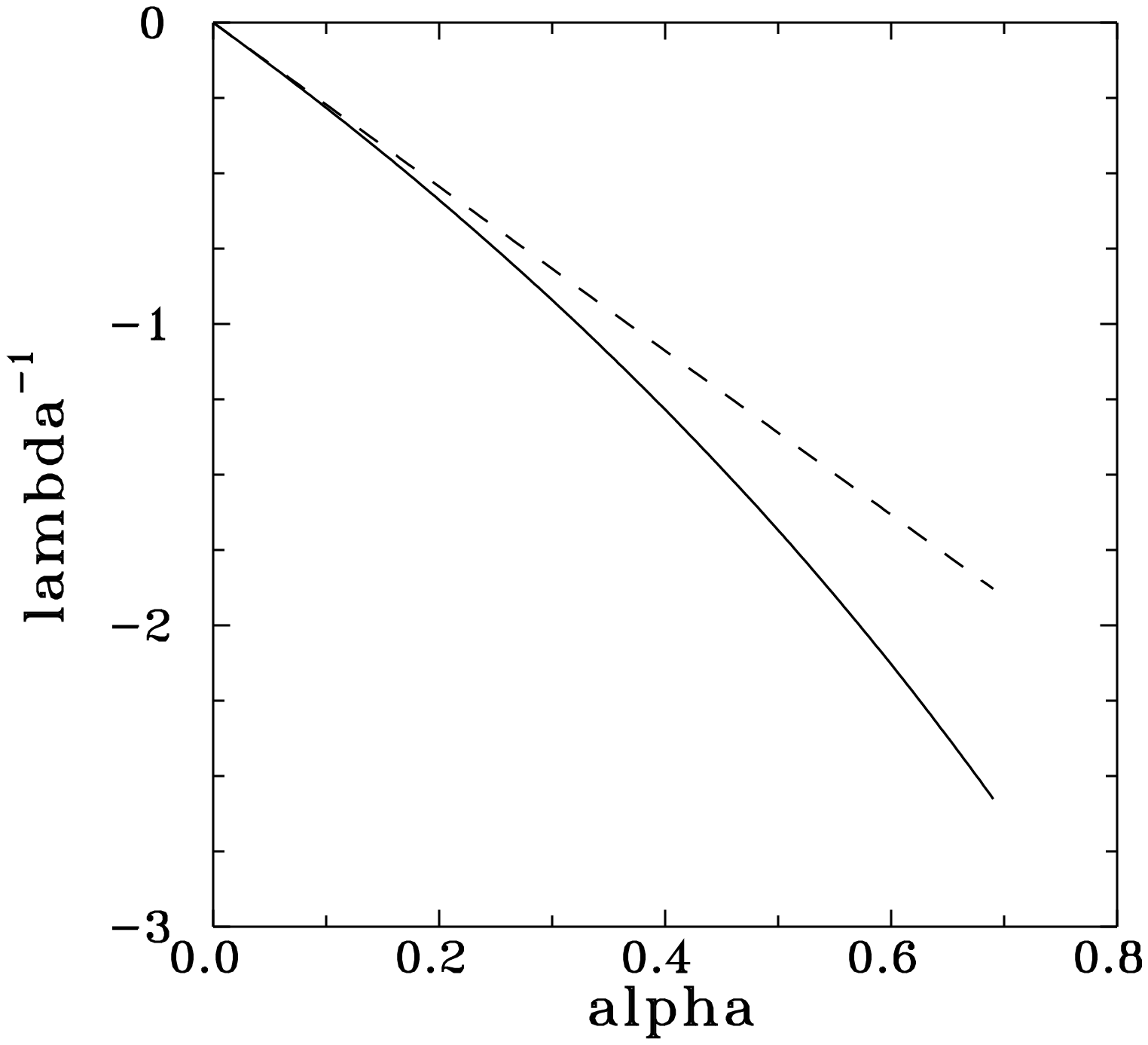}{0.8}    %**epsf** 
  \vskip -1cm 
Inverse renormalized strength $\lambda^{-1}_{\cal R}(m_\pi^2)$ in units of 
$m_r$ as a  function
of the Coulomb intensity potential parameter $\alpha$ for a 
pion mass of 140 MeV. Nonperturbative calculation (solid line) and
first-order perturbative calculation (dashed line), see text 
for the explanation.

%%%%%%%%%%%%%%%%%%%%%%%%%%%%%% FIG. 6 %%%%%%%%%%%%%%%%%%%%%%%%%%%%%%%%%%%%%% 
\vspace{1 true cm} 
\item[Fig. 6]  
\postscript{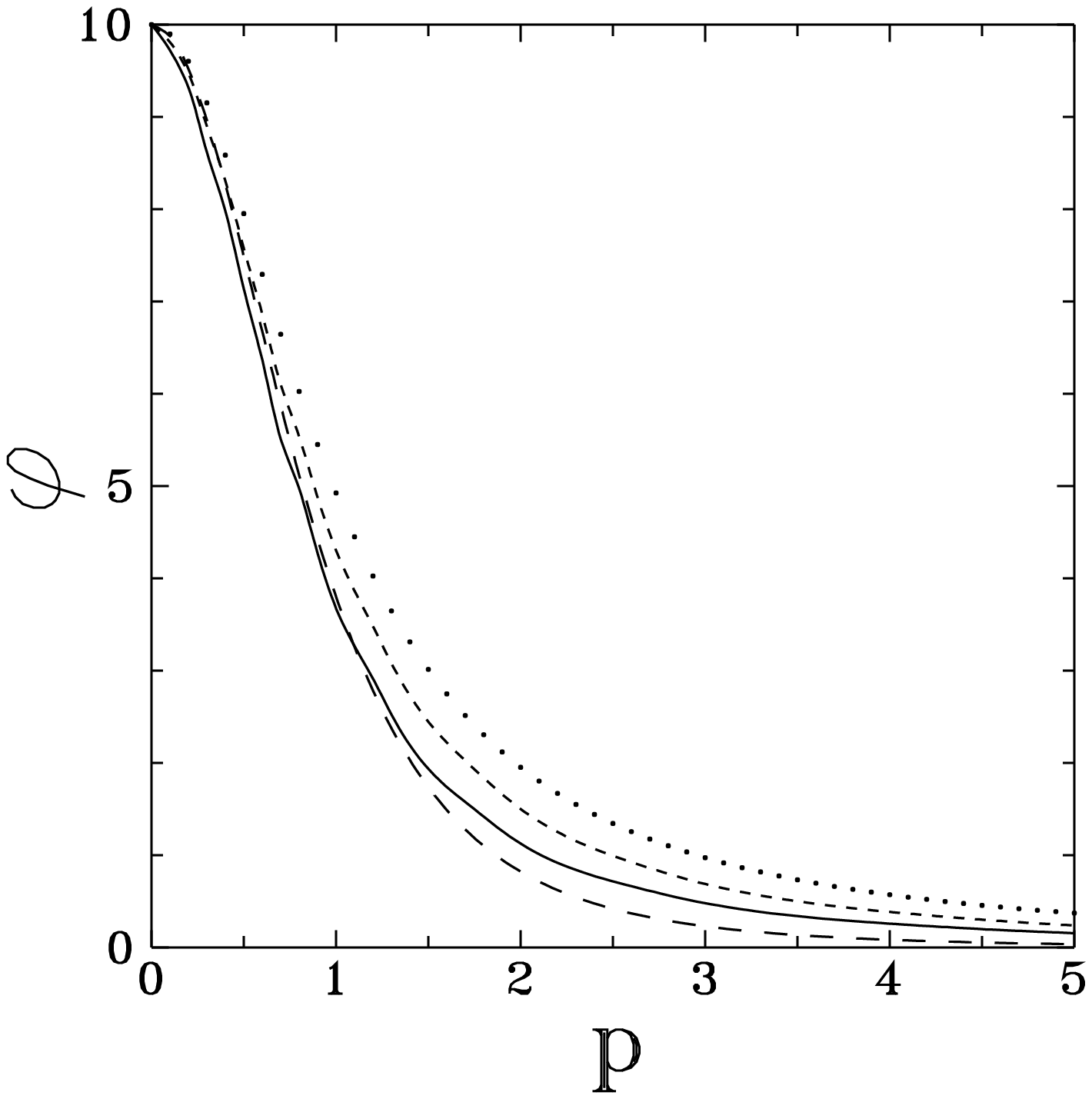}{0.8}    %**epsf** 
 \vskip -1cm 
Pion wave-function ($\varphi$) with arbitrary normalization 
as a function of momentum in units of quark mass. Calculations performed 
with $A(k)=1$. Pion model wave-function model of (\ref{piphip})
for $\alpha=0.6904$ and $m=406$ MeV (solid line), first-order 
perturbative calculation from Eq.(\ref{piphipert}) (short-dashed line),
asymptotic form Eq. (\ref{piphi0}) (dotted line) and fit 
from Eq.(\ref{piphia}) (dashed line).

%%%%%%%%%%%%%%%%%%%%%%%%%%%%%% FIG. 7 %%%%%%%%%%%%%%%%%%%%%%%%%%%%%%%%%%%%%% 
\vspace{1 true cm} 
\item[Fig. 7]  
\postscript{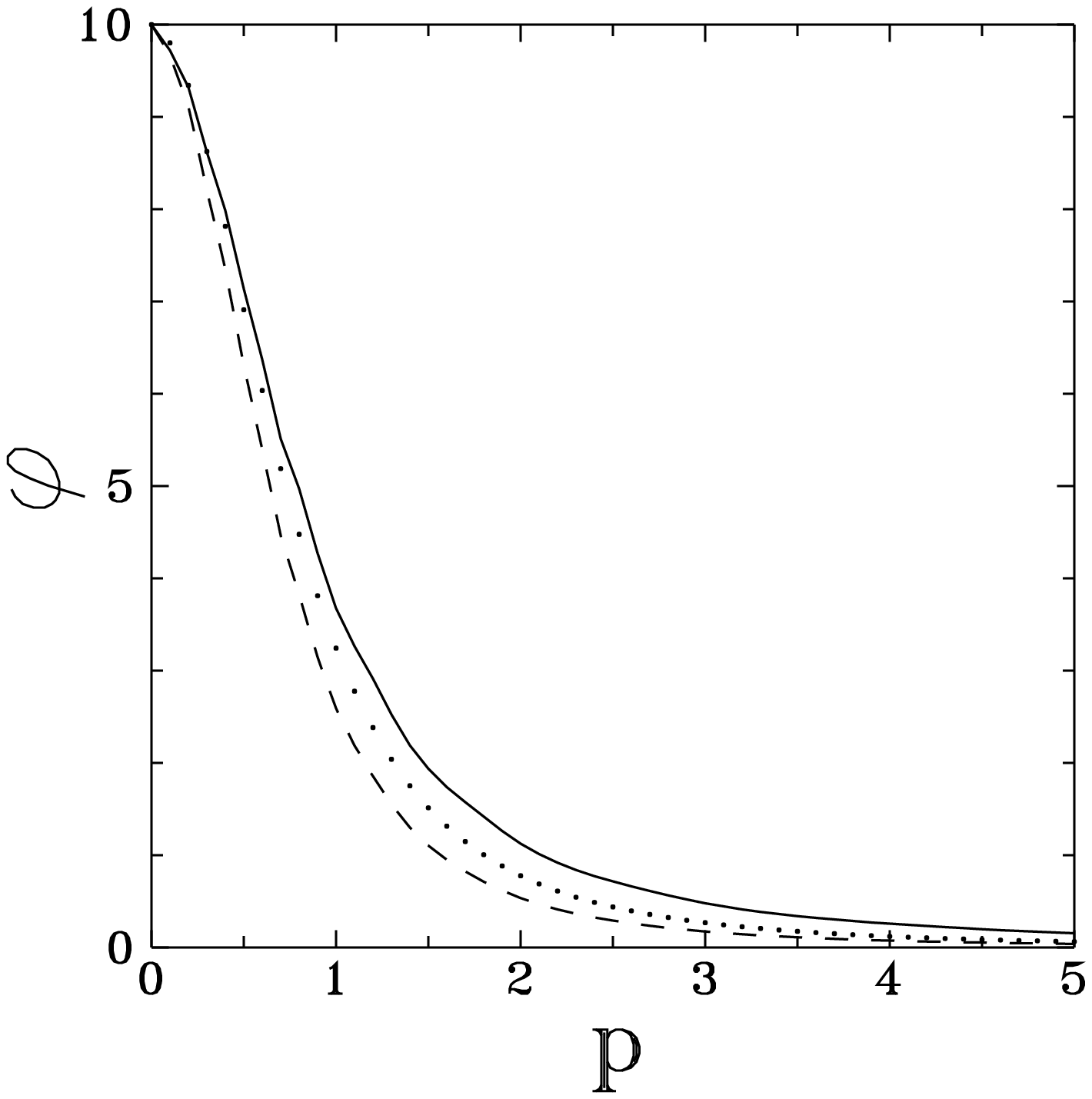}{0.8}    %**epsf** 
 \vskip -1cm 
Pion wave-function ($\varphi$) as a function of momentum in units of 
quark mass. Pion model for $\alpha=0.6904$ and $m=406$ MeV, 
calculation with Eq.(\ref{piphip}) (dashed line); considering $A(k)=1$ 
(solid line). Results for $\alpha=0.18$ and $m=386$ MeV (dotted line).

%%%%%%%%%%%%%%%%%%%%%%%%%%%%%% FIG. 8 %%%%%%%%%%%%%%%%%%%%%%%%%%%%%%%%%%%%%% 
\vspace{1 true cm} 
\item[Fig. 8]  
\postscript{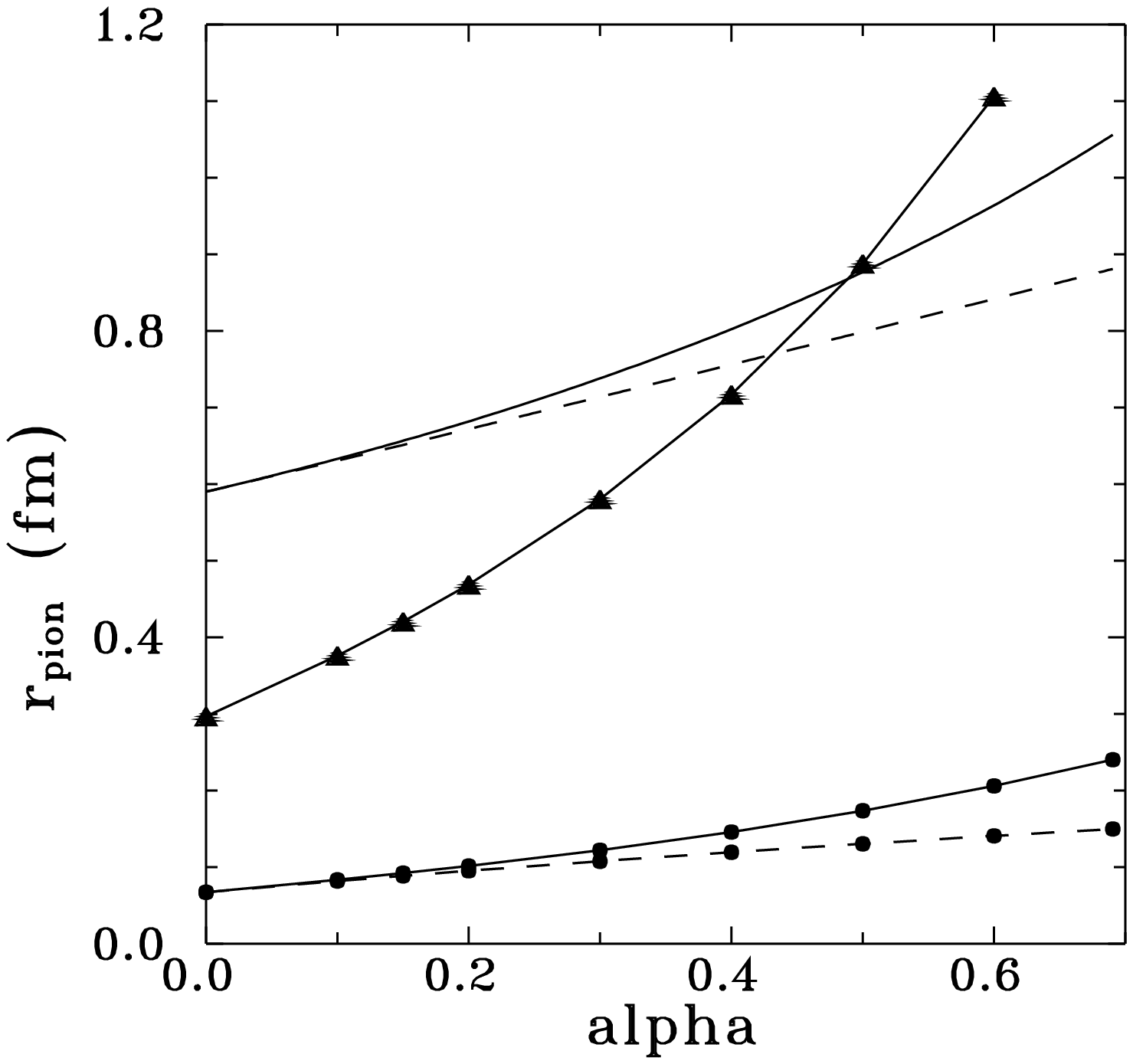}{0.8}    %**epsf** 
  \vskip -1cm 
Pion charge radius  as a  function
of the strength $\alpha$ of the Coulomb potential. Quark mass
of 406 MeV. Results for Eq.(\ref{ffactor}): nonperturbative calculation 
of the wave function (solid line) and first order
perturbative calculation (dashed line). 
Results for Eq.(\ref{wsffactor}) and wave-function obtained
with $A(k)=1$ (solid line with triangles). Results for 
Eq. (\ref{nwsffactor}) with the wave function obtained with
$A(k)=1$: nonperturbative calculation 
of the wave function (solid line with dots) and first order
perturbative calculation (dashed line with dots). 

%%%%%%%%%%%%%%%%%%%%%%%%%%%%%% FIG. 9 %%%%%%%%%%%%%%%%%%%%%%%%%%%%%%%%%%%%%% 
\vspace{1 true cm} 
\item[Fig. 9]  
\postscript{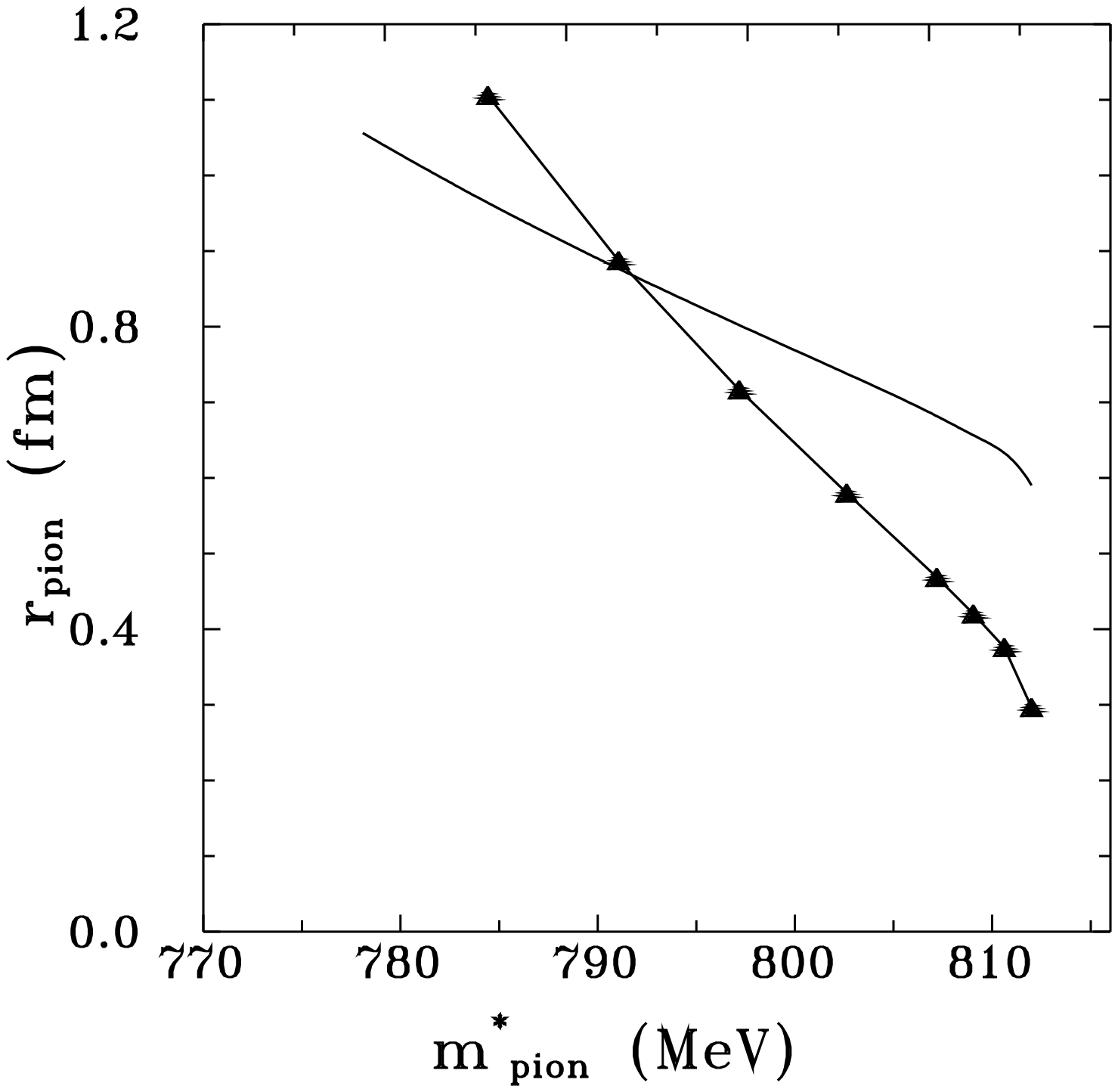}{0.8}    %**epsf** 
  \vskip -1cm 
Pion charge radius  as a  function
of the mass of the excited $q\overline q$ state for a constrained
pion mass of 140 MeV. The strength $\alpha$ is within the range of Fig.6
and the quark mass is 406 MeV.
Results for Eq.(\ref{ffactor}) (solid line) and for 
Eq.(\ref{wsffactor}) with wave-function obtained
considering $A(k)=1$ (solid line with triangles).

%%%%%%%%%%%%%%%%%%%%%%%%%%%%%% FIG. 10 %%%%%%%%%%%%%%%%%%%%%%%%%%%%%%%%%%%%%% 
\vspace{1 true cm} 
\item[Fig. 10]  
\postscript{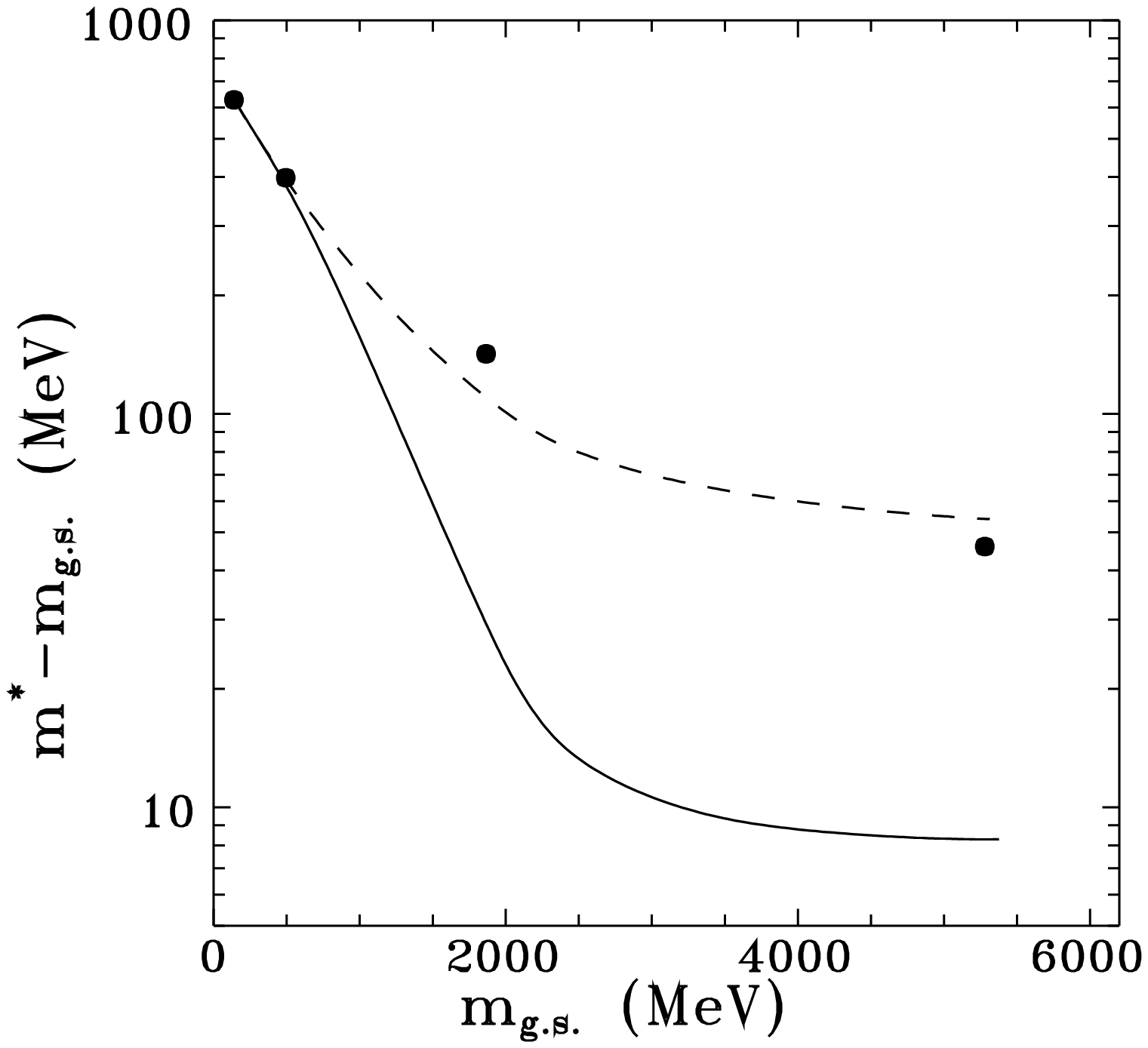}{0.8}    %**epsf** 
 \vskip -1cm 
The difference between the masses of the excited and ground state of
the $\overline q q$ system as a funtion of the ground state mass in
the effective relativistic model. The quark  mass is varied in the range
between 386 and 5000 MeV. The other quark mass is fixed to 386 MeV. 
Calculations with $\alpha=0.18$ (solid line) and $\alpha=0.5$ (dashed line).
The experimental values from Table II are given by the full circles.
\end{description} 
}}


\begin{thebibliography}{99}

\bibitem{pauli0}S. J. Brodsky, H.C. Pauli, and S.S. Pinsky,
Phys. Rep. {\bf 301} (1998) 299.

\bibitem{pauli2} H.C. Pauli, 
%" On the effective light-cone QCD-Hamiltonian:
%Application to the pion and other mesons", 
Nucl. Phys. {\bf B} (Proc. Supp.) {\bf 90} (2000) 154.

\bibitem{pauli1} H.C. Pauli, %"A Compedium of Light-Cone Quantization ",
Nucl. Phys. {\bf B} (Proc. Supp.) {\bf 90} (2000) 259.

\bibitem{pauli3} H.C. Pauli, Eur. Phys. J. {\bf C7} (1998) 289.
\bibitem{pauli4} H.C. Pauli,
"DLCQ and the effective interactions in hadrons" in: New Directions
in Quantum Chromodynamics, C.R. Ji and D.P. Min, Editors, American
Institute of Physics, 1999, p. 80-139.

\bibitem{saw1} M. Sawicki, Phys. Rev. {\bf D32} (1985) 2666.

\bibitem{t1} T.Frederico, V.S. Tim\'oteo, and L.Tomio, 
Nucl.Phys. {\bf A653} (1999) 209.

\bibitem{t2} T. Frederico, A. Delfino, and  L. Tomio,
Phys. Lett. B {\bf 481} (2000) 143.

\bibitem{t3} T. Frederico, A.Delfino, L.Tomio, and V.S. Tim\'oteo,
"Fixed Point Hamiltonians in Quantum Mechanics", hep-ph/0101065.

\bibitem{biel} S. Bielefeld, J. Ihmels, and  H.C. Pauli,
"On technically solving an effective light-cone Hamiltonian",
hep-th/9940241.

\bibitem{izuber} C. Itzykson and J. B. Zuber, "Quantum Field Theory";
McGraw-Hill, 1980.

\bibitem{mill} T. Frederico and G. A. Miller, 
Phys. Rev. {\bf D45} (1992) 4207.

\bibitem{pach}J.P.B.C de Melo, H.W.Naus, and T.Frederico, 
Phys. Rev. {\bf C59} (1999) 2278.

\bibitem{tarr} R. Tarrach, Z. Phys. {\bf C2} (1979) 221.

\bibitem{amen} S.R. Amendolia et al., Phys. Lett. {\bf B178} (1986) 116.
\end{thebibliography}
\end{document}